\documentclass[a4paper,11pt,preprintnumbers,letterpaper,amsmath,amssymb,nofootinbib]{revtex4-2}

\usepackage[dvipdfmx]{graphics}
\usepackage{mathtools}
\usepackage[T1]{fontenc}
\usepackage{braket}	
\usepackage{dcolumn}
\usepackage{comment}
\usepackage{url}
\usepackage{hyperref}
\usepackage{MnSymbol}

\newcommand{\LQ}{\Lambda_{\rm QCD}}
\newcommand{\alfs}{\alpha_{s}}
\newcommand{\msbar}{\overline{\rm MS}}
\newcommand{\mbar}{\overline{m}}
\newcommand{\LMS}{\Lambda_{\rm \overline{MS}}}

\newcommand{\bea}{\begin{eqnarray}}
\newcommand{\eea}{\end{eqnarray}}
\newcommand{\simgt}{\hbox{ \raise3pt\hbox to 0pt{$>$}\raise-3pt\hbox{$\sim$} }}
\newcommand{\simlt}{\hbox{ \raise3pt\hbox to 0pt{$<$}\raise-3pt\hbox{$\sim$} }}

\newcommand{\be}{\begin{equation}}
\newcommand{\ee}{\end{equation}}

\newcommand{\non}{\nonumber \\}

\begin{document}

\preprint{TU-1183, KEK-TH-2510}

\title{\boldmath
Inclusive $|V_{cb}|$ determination in $\overline{\mathrm{MS}}$ mass scheme
using dual-space-renormalon-subtraction method}

\author{Yuuki Hayashi}
\email{yuuki.hayashi.s3@alumni.tohoku.ac.jp}
\affiliation{Department of Physics, Tohoku University,
Sendai, 980-8578, Japan}
\author{Go Mishima}
\email{go.mishima@icloud.com}
\affiliation{Department of Physics, Tohoku University,
Sendai, 980-8578, Japan}
\author{Yukinari Sumino}
\email{yukinari.sumino.a4@tohoku.ac.jp}
\affiliation{Department of Physics, Tohoku University,
Sendai, 980-8578, Japan}
\author{Hiromasa Takaura}
\email{hiromasa.takaura@yukawa.kyoto-u.ac.jp}
\affiliation{Institute of Particle and Nuclear studies, KEK,
Tsukuba, 305-0801, Japan}

\begin{abstract}
We determine $|V_{cb}|$
from the inclusive semileptonic decay width
of the $B$ meson
with the known N$^3$LO
perturbative coefficients
for the first time in the $\overline{\mathrm{MS}}$ mass scheme.
We make use of a recently 
developed method,
dual-space-renormalon-subtraction (DSRS) method,
to separate and subtract the order
$\Lambda_{\rm QCD}^2/m_b$
infrared renormalon.
This allows us to
perform the analysis accurately in the
$\overline{\mathrm{MS}}$ mass scheme,
in which otherwise the perturbative series
does not converge well up to the currently
calculated perturbation order.
Our result reads $|V_{cb}|=0.0415\,(^{+10}_{-12})$,
which is consistent with 
the previous results based on other mass schemes
and showing an independent cross check
of the current theoretical evaluation
of the inclusive decay width.
\end{abstract}

\maketitle

\section{Introduction}
\label{chap1}
The Cabibbo-Kobayashi-Maskawa matrix elements play key roles 
in studying
flavor physics such as $CP$ violation in the Standard Model (SM)
of particle physics,
and 
one of the elements, $|V_{cb}|$, is determined 
from 
the semileptonic $B$ decay processes $B\to X_c\ell \nu$.
It is known, however,
that there is a sizable tension between
the $|V_{cb}|$ values 
determined from the exclusive decays $\bar{B}\to D^{(*)}\ell \nu$
and from the inclusive decays $B\to X_c\ell \nu$.
This is called the $|V_{cb}|$ puzzle.
According to the Particle Data Group (PDG) \cite{ParticleDataGroup:2022pth}, 
the average values of the exclusive and inclusive analyses are given,
respectively, by 
\be
|V_{cb}|_{\rm excl.}=0.0394(8),~~~~
|V_{cb}|_{\rm incl.}=0.0422(8).
\label{Vcb-PDG}
\ee
It is argued that 
the $|V_{cb}|$ puzzle
is difficult to be explained by physics beyond the SM
\cite{Crivellin:2014zpa}.
Under this circumstance,
reconfirmation of 
the SM analysis
is important.

In the theoretical calculation of the inclusive decays, 
the operator product expansion (OPE) is used.
It provides a theoretical prediction of the decay width 
$\Gamma(B\to X_c\ell \nu )$ 
in the double expansion in the strong coupling constant $\alpha_s$ 
and inverse of the bottom quark mass $1/m_b$.
In the first step of constructing the OPE of $\Gamma$,
the pole mass scheme for the bottom quark
is used because 
it is a natural scheme in implementing nonrelativistic dynamics of the bottom quark.
At this point,
the infrared (IR) renormalons are present 
in the perturbative series expression of the leading order (LO) Wilson coefficient  (i.e., the leading term in the OPE),
and the perturbative series exhibits a factorial divergence.
They induce 
${\cal O}((\LQ/m_b)^n)$ uncertainties ($\LQ\sim300~$MeV), 
obscuring systematic improvement of theoretical accuracy by the double expansion.
Switching from the pole mass scheme 
to a short-distance mass scheme,
one can eliminate the $u=1/2$ renormalon (${\cal O}(\LQ/m_b)$ uncertainty)
in the LO Wilson coefficient,
resulting in a better convergence.
Recent developments in computational techniques 
have accomplished the
next-to-next-to-next-to-LO (N$^3$LO) calculations of 
the inclusive decay width \cite{Fael:2020tow} and 
the lepton energy moments in the $b\to c\ell \nu$ 
decay~\cite{Fael:2022frj}.
These lead to the latest determinations of
$|V_{cb}|$,
which are in agreement with the current PDG inclusive value:
Refs.~\cite{Alberti:2014yda,Bordone:2021oof,Bernlochner:2022ucr}
adopt the kinetic mass scheme~\cite{Bigi:1994ga,Bigi:1996si}, 
and Refs.~\cite{Hayashi:2022hjk,HeavyFlavorAveragingGroup:2022wzx}
adopt the 1S mass scheme~\cite{Hoang:1998hm,Hoang:1998ng,Bauer:2004ve}.

In this paper,
we report a $|V_{cb}|$ determination from
$\Gamma(B\to X_c\ell \nu )$ 
with the N$^3$LO result adopting 
the $\msbar$ mass scheme for the first time.
It was hitherto missing 
because the convergence of the perturbative series in terms of the $\msbar$ mass 
is known to be slow
even after the $u=1/2$ renormalon is canceled,
owing to
the fact that the $\msbar$ mass ($\approx4.2$ GeV) 
is far from the pole mass ($\approx 4.8$ GeV).
In particular,
$\Gamma$ is proportional to $m_b^5$, and thus
the slow convergence of the mass relation seriously affects the convergence of $\Gamma$.
Nevertheless, since the decay width does not depend on the choice of short-distance
masses, the prediction of $|V_{cb}|$ in the $\msbar$ mass scheme should match the prediction in other mass
schemes (as more terms of the perturbation series are included). This is an important cross-check
of the inclusive determinations of $|V_{cb}|$.

We use the dual-space-renormalon-subtraction (DSRS) method~\cite{Hayashi:2023fgl}
and subtract the IR renormalon at $u=1$, i.e., the ${\cal O}((\LQ/m_b)^2)$ uncertainty, 
aiming at better accuracy in the determination in the $\msbar$ mass scheme.
The DSRS method is founded on the well-known Borel method~\cite{Beneke:1998ui},
which provides a regularization
of a divergent series caused by renormalons.
The major drawback of the Borel method 
is that it requires the information of all-order perturbative series
to construct the Borel transform,
which is impossible in practical application.
The DSRS method 
enables us
to construct an approximate 
Borel transform
from a finite order of perturbative series,
and to obtain a converging result  
as we increase the perturbative order.

We explain the outline of the DSRS method~\cite{Hayashi:2023fgl}.
Given an observable whose typical scale is $Q \gg \LQ$, we perform a dual transform using the inverse Laplace integral
to give a quantity which now depends on the dual-space variable $t$ instead of $Q$.
The IR renormalons are suppressed in the dual-space quantity 
due to a property of the inverse Laplace transform; see, for instance, Eq.~\eqref{mass-renormalon-suppression} below.
In particular, when a renormalon uncertainty is given by the integer power in $\LQ/Q$,
the renormalon can be completely eliminated by the (simplest formula of) DSRS method.
However, the simplest formula is not sufficient to completely eliminate
renormalon uncertainties in the case that they deviate from the integer power in $\LQ/Q$;
the general form of renormalon uncertainties is given by $f(\alpha_s(Q^2)) (\LQ/Q)^n$,
where $n$ is an integer and $f(\alpha_s(Q^2))$ can be calculated perturbatively in the form,
$\alpha_s(Q^2)^{\tilde{\nu}} \sum_{n=0}^{\infty} c_n \alpha_s(Q^2)^n$ ($\tilde{\nu}$ can be non-integer),
a function of $\LQ/Q$.
There is an extended formula which can eliminate general renormalon uncertainties
but it complicates calculations.
For this reason, in this paper, we use the simplest formula for simplicity of our analysis.
We note, however, that the relevant renormalon uncertainties are largely removed because
$f(\alpha_s(Q^2))$ is given by $f(\alpha_s(Q^2))=const.+\mathcal{O}(\alpha_s(Q^2))$ in our study.
The residual renormalon uncertainties are considered in our analyses of systematic uncertainties.


The procedure for our $|V_{cb}|$ determination is as follows. 
As a first trial, we treat only the OPE of the decay width $\Gamma(B\to X_c\ell \nu)$ and not those of the moments.
After rewriting the OPE in the $\msbar$ mass, we apply the DSRS method to the LO Wilson coefficient of $\Gamma$ and then the imaginary part caused by
the $u=1$ renormalon is separated. 
The separated $u=1$ renormalon is assumed to be absorbed and removed by the first non-perturbative term of the OPE, the kinetic energy term of the $B$ meson, $\mu_{\pi}^2$. 
In this study, the non-perturbative matrix elements, $\mu_{\pi}^2$ (and $\mu_G^2$), 
are determined from the $B$ and $D$ meson masses after the removal of renormalons in the $1/m_h$ expansion
for the heavy-light meson mass, by using the DSRS method.
Following this procedure we can obtain the OPE for $\Gamma$ where the non-perturbative effects are incorporated. 
We note that our double expansion is systematic because of a proper treatment of renormalons.
By comparing our renormalon-subtracted OPE with an experimental value of $\Gamma$, 
we determine the inclusive $|V_{cb}|$ in the $\msbar$ mass scheme and compare it with previously determined values in other short-distance mass schemes.

The paper is organized as follows.
Section~\ref{sec2}
is devoted to an explanation of the theoretical
framework of our analysis,
namely the OPE and the renormalon subtraction.
In Section~\ref{sec3},
we examine renormalons of the pole masses to determine the non-perturbative matrix elements
needed in our determination of $|V_{cb}|$.
Section~\ref{sec4}
is the main part,
where $|V_{cb}|$
is determined
and compared to the previous studies.
We summarize and conclude our study 
in Section~\ref{chap5}.
For the readers' convenience,
we collect the formulas from the literature 
used in our analysis in Appendix~\ref{App.A}.
In Appendix~\ref{App.B}, we explain our method to estimate the higher-order
perturbative coefficient for the relation between the pole and the $\msbar$ masses.

\section{\boldmath OPE in $\msbar$ mass scheme and renormalon subtraction}
\label{sec2}
\subsection{\boldmath OPE of the decay width in $\msbar$ mass scheme}
\label{sec2.1}
In the framework of the OPE based on 
the heavy quark 
effective theory (HQET) \cite{Georgi:1990um,Eichten:1989zv,Grinstein:1990mj}, 
the inclusive semileptonic decay width 
$\Gamma(B\to X_c\ell \nu)$ 
is given by
\be
\Gamma=\frac{G_F^2|V_{cb}|^2}{192\pi^3}A_{ew}m_b^5\bigg[C_{\bar{Q}Q}^\Gamma(m_b,\rho)+C_{\rm kin}^\Gamma(m_b,\rho)\frac{\mu_\pi^2}{m_b^2}+C_{cm}^\Gamma(m_b,\rho)\frac{\mu_G^2(m_b)}{m_b^2}+{\cal O}\bigg(\frac{\LMS^3}{m_b^3}\bigg)\bigg].
\label{OPE-Gamma-pole}
\ee
Here, $G_F\approx 1.166 \times 10^{-5} \,\rm GeV^{-2}$ is the Fermi constant 
and $A_{ew}\approx1.014$ is the electroweak correction to the decay width. 
The uncertainties of $G_F$ and $A_{ew}$ are negligible in this analysis.
$\LMS$ is the QCD scale parameter in the $\msbar$ scheme.
$m_b$ $(m_c)$ denotes 
the pole mass of the bottom (charm) quark and $\rho=m_c/m_b$.
$\mu_\pi^2$ and $\mu_G^2$ are non-perturbative matrix elements in HQET given by 
\be
\mu_\pi^2=-{\langle H(v)|\bar{h}_v(i{D}_\perp)^2 h_v|H(v)\rangle},~~~~
\mu_G^2=-{\langle H(v)|\bar{h}_v\frac{g_s}{2}\sigma_{\mu\nu}G^{\mu\nu} h_v|H(v)\rangle},
\label{mupimuG}
\ee
where $D^\mu$ is the covariant derivative, $G^{\mu\nu}$ is the field strength tensor, $\sigma^{\mu\nu}=\frac{i}{2}[\gamma^\mu,\gamma^\nu]$ and $D_\perp^\mu=D^\mu-v^\mu(v\!\cdot\!D)$.
In Eq.~\eqref{mupimuG}, $h_v$ is the heavy quark field in the HQET, $v=p_B/M_B$ is the velocity of the
$B$ meson and $|H(v)\rangle$ is the meson state in the infinite mass limit.
Note that the above matrix elements are identical for $H=B,\,D$ (or $h=b,\,c$).
$C_{\bar{Q}Q}^\Gamma,\,C_{\rm kin}^\Gamma$ and $C_{\rm cm}^\Gamma$ are Wilson coefficients which are calculated perturbatively.
Up to ${\cal O}(1/m_b^2)$, $C_{\rm kin}^\Gamma=-\frac{1}{2}C_{\bar{Q}Q}^\Gamma$ due to the reparameterization invariance (RPI) \cite{Luke:1992cs,Mannel:2018mqv}.\footnote{
Reparameterization invariance is due to the Lorentz invariance of 
full QCD, which fixes the renormalization constant of the kinetic energy term $\mu_\pi^2$.}
At present, $C_{\bar{Q}Q}^\Gamma=\sum_{n=0}^\infty X_n(\rho)\alpha_s(m_b)^{n}$ and $C_{cm}^\Gamma=\sum_{n=0}^\infty Y_n(\rho)\alpha_s(m_b)^{n}$ are calculated up to ${\cal O}(\alfs^3)$ \cite{Luke:1994yc,Trott:2004xc,Aquila:2005hq,Pak:2008qt,Pak:2008cp,Melnikov:2008qs,Dowling:2008mc,Fael:2020tow} and ${\cal O}(\alfs)$ \cite{Bigi:1992su,Blok:1993va,Manohar:1993qn,Alberti:2013kxa}, respectively.
Note that ${\cal O}(\LMS)$ non-perturbative terms such as $\bar{\Lambda}$ do not appear for
this observable because the HQET Lagrangian does not contain the dimension-four operator whose form is $\bar{h}_v\,{\cal O}\,h_v$ (up to the equation of motion).

We rewrite the OPE by the $\msbar$ mass $\mbar_h$, using the pole-$\msbar$ mass relation 
up to ${\cal O}(\alpha_s^3)$~\cite{Melnikov:2000qh,Marquard:2007uj,Chetyrkin:1999ys,Chetyrkin:1999qi}.
Then the $u=1/2$ renormalon of the LO Wilson coefficient $C_{\bar{Q}Q}^\Gamma$ is canceled with that of the prefactor $m_b^5$ in Eq.~\eqref{OPE-Gamma-pole}.
The expression is given by
\be
\Gamma=\frac{G_F^2|V_{cb}|^2}{192\pi^3}A_{ew}\mbar_b^5\bigg[\bar{C}_{\bar{Q}Q}^\Gamma(\mbar_b, \bar{\rho}) \Big(1-\frac{\mu_\pi^2}{2m_b^2}\Big)+\bar{C}_{cm}^\Gamma(\mbar_b, \bar{\rho}) \frac{\mu_G^2}{m_b^2}+{\cal O}\bigg(\frac{\LMS^3}{m_b^3}\bigg)\bigg],
\label{OPE-Gamma-mbar}
\ee
where the expansion parameter is kept to be the inverse of the pole mass.
$\bar{C}_{\bar{Q}Q}^\Gamma\equiv(m_b/\mbar_b)^5{C}_{\bar{Q}Q}^\Gamma$ and $\bar{C}_{cm}^\Gamma\equiv(m_b/\mbar_b)^5{C}_{cm}^\Gamma$ are calculated 
by rewriting $m_b$ and $m_c$ in terms of $\mbar_b$ and $\mbar_c$.
We define $\bar{\rho}=\mbar_c/\mbar_b$.
The LO renormalon of $\bar{C}_{\bar{Q}Q}^\Gamma$ is at $u=1$,\footnote{
In the large-$\beta_0$ approximation, the $u=1$ renormalon is absent both in the pole-$\msbar$ mass relation and in $C_{\bar{Q}Q}^\Gamma$.
Beyond this approximation, 
it is possible that the $u=1$ renormalon exists and therefore we assume 
that it does.
} generating the imaginary part of ${\cal O}(\LMS^2/m_b^2)$, if $\bar{C}_{\bar{Q}Q}^\Gamma$ is regularized in the Borel resummation method.
The Borel-resummation regularization of $\bar{C}_{\bar{Q}Q}^\Gamma$ is given by
\be
\big[\bar{C}_{\bar{Q}Q}^\Gamma\big]_\pm=\big[\bar{C}_{\bar{Q}Q}^\Gamma\big]_{\rm PV}\pm i\,{\rm Im}\,\big[\bar{C}_{\bar{Q}Q}^\Gamma\big],
\label{barCpm-def}
\ee
where $\big[\bar{C}_{\bar{Q}Q}^\Gamma\big]_{\rm PV}$ is the regularized Wilson coefficient defined by the principal value integral of the Borel resummation integral and 
\be
{\rm Im}\,\big[\bar{C}_{\bar{Q}Q}^\Gamma\big]={\cal O}(\LMS^2/m_b^2).
\ee
(See, e.g., Sec.~3.1 of Ref.~\cite{Hayashi:2023fgl} for a review of the Borel resummation method.)
Identification of a renormalon-free contribution in the Wilson coefficient is scheme dependent, 
and the conventional way is to identify it with the principal value (PV) part in Eq.~\eqref{barCpm-def}.
The imaginary part of ${\cal O}(\LMS^2/m_b^2)$ 
is expected to be absorbed by the non-perturbative part $\mu_\pi^2$,
realizing the cancellation of the IR renormalon.
In this way, the OPE in the double expansion in $\alpha_s$ and $1/m_b$
can achieve an accurate description of $\Gamma$,
and in principle enables precise determination of $|V_{cb}|$
in the $\msbar$ mass scheme.

\subsection{Non-perturbative matrix elements}
\label{sec2.2}
In this subsection 
we summarize 
the property of 
the non-perturbative matrix elements $\mu_\pi^2$ and $\mu_G^2$,
which is the prerequisite of the analysis in the $1/m_h$ expansion of the $B$ and $D$ meson masses.
The masses of $H=B$ and $D$ are formulated  by 
\be
M_H^{(s)}=m_h+\bar{\Lambda}+\frac{\mu_\pi^2}{2m_h}+A(s)C_{cm}^M(m_h)\frac{\mu_G^2(m_h)}{2m_h}+{\cal O}\bigg(\frac{\LMS^3}{m_h^2}\bigg),
\ee
where $s\,(=0,\,1)$ denotes the spin of $H$.
The leading contribution to $M_H$ is the heavy quark pole mass $m_h$ $(h=b,\,c)$,
which has an uncertainty due to the IR renormalons starting from ${\cal O}(\LMS)$.
The leading non-perturbative correction is ${\cal O}(\LMS)$,
and in the sum, namely in $m_h+\bar{\Lambda}$, there is no uncertainty of ${\cal O}(\LMS)$.
The first non-perturbative correction $\bar{\Lambda}={\cal O}(\LMS)$ 
is the contribution from the light degrees of freedom of $H$, 
which can be written by the matrix element of the light sector of 
the QCD Hamiltonian.
By construction of the effective field theory,
$\mu_\pi^2$ and $\mu_G^2$ are identical to those in Eq.~\eqref{mupimuG}.
The Wilson coefficient of $\mu_\pi^2/(2m_h)$ is exactly 1 due to RPI,
and $C_{cm}^M$ has been calculated up to ${\cal O}(\alpha_s^3)$ \cite{Grozin:2007fh}.
Since the chromomagnetic interaction breaks the spin symmetry in the HQET Lagrangian, 
the $\mu_G^2$ term is proportional to a spin-dependent coefficient $A(s)$; $A(s)=-1$ for $s=0$ (pseudo-scalar meson $H^{*}$) and $A(s)=1/3$ for $s=1$ (vector meson $H$).

Up to this order, we can separate the spin-dependent part by a simple combination of $M_H^{(s)}$.
A clever choice is to express it by the difference of the mass squared 
of the $H$ mesons as
\be
\frac{3}{4}\big(M_{H^*}^2-M_H^2\big)
=C_{cm}^M(m_h)\mu_G^2(m_h)
+{\cal O}\bigg(\frac{\LMS^3}{m_h}\bigg),
\label{hyp-fine}
\ee
where power dependence on the pole mass $m_h$ does not appear in the LO calculation.
From this $1/m_h$ expansion, one can see that 
$\mu_G^2$ does not have an ${\cal O}(\LMS^2)$ renormalon (at $u=1$).
(If it had, there were no quantities which could cancel it.)
Thus, we do not consider the IR renormalon of this part in the following analysis.
In Ref.~\cite{Hayashi:2022hjk}, the value of $\mu_G^2(\mbar_b)$ using $C_{cm}^M$ at N$^3$LO \cite{Grozin:2007fh} is determined as
\be
\mu_G^2(\mbar_b)=0.284\pm0.014~{\rm GeV^2},
\label{res-muG2}
\ee
in which the uncertainty is comparable to ${\cal O}(\LMS^3/\mbar_b)$, the power correction in Eq.~\eqref{hyp-fine},
reflecting the fact that  the fixed-order result of $C_{cm}^M$ contains the $u=1/2$ renormalon.\footnote{
Although we neglect the power correction indicated in Eq.~\eqref{hyp-fine},
its typical uncertainty is properly reflected in Eq.~\eqref{res-muG2} in this manner.}

The $\mu_G^2$-independent part is given by a linear combination of $M_H$ and $M_{H^*}$ as
\be
\big[M_H\big]_{\rm spin\,ave.}=\frac{M_H+3M_{H^*}}{4}=m_h+\bar{\Lambda}+\frac{\mu_\pi^2}{2m_h}+{\cal O}\bigg(\frac{\LMS^3}{m_h^2}\bigg).
\label{OPE-MH}
\ee
We assume that the largest two IR renormalon contributions 
are those corresponding to $u=1/2$ and $u=1$.
The imaginary part from those IR renormalons are absorbed into $\bar{\Lambda}$ and $\mu_\pi^2$, respectively.
The imaginary part of the pole mass in the PV prescription is expressed as
\be
\big[m_h\big]_\pm=\big[m_h\big]_{\rm PV}\pm iN_{1/2}\LMS\pm iN_{1}\frac{\LMS^2}{\big[m_h\big]_{\rm PV}}+{\cal O}\bigg(\frac{\LMS^3}{\big[m_h\big]_{\rm PV}^2}\bigg),
\label{mhpm}
\ee
where $\big[m_h\big]_\pm$ is the regularized pole mass. 
$\big[m_h\big]_{\rm PV}$ denotes the renormalon-subtracted pole mass in the PV prescription 
and we call it the PV mass of $h$.
$N_{1/2}$ and $N_1$ are the normalization constants of the imaginary part.
We note that the Wilson coefficients for $\bar{\Lambda}$ and $\mu_\pi^2$ are 
independent of $m_b$.
The renormalon-subtracted formula of $\big[M_H\big]_{\rm spin\,ave.}$ is expressed by
\be
\big[M_H\big]_{\rm spin\,ave.}=\big[m_h\big]_{\rm PV}+\big[\bar{\Lambda}\big]_{\rm PV}+\frac{\big[\mu_\pi^2\big]_{\rm PV}}{2\big[m_h\big]_{\rm PV}}+{\cal O}\bigg(\frac{\LMS^3}{m_h^2}\bigg),
\label{MH-ren-sub-OPE}
\ee
where $\big[\bar{\Lambda}\big]_{\rm PV}$ and $\big[\mu_\pi^2\big]_{\rm PV}$ are defined by
$\big[\bar{\Lambda}\big]_{\rm PV}=\bar{\Lambda}\pm iN_{1/2}\LMS$
and
$\big[\mu_\pi^2\big]_{\rm PV}=\mu_\pi^2\pm iN_{1}{\LMS^2}$.

The matrix elements $\big[\bar{\Lambda}\big]_{\rm PV}$ and $\big[\mu_\pi^2\big]_{\rm PV}$ can be determined by comparing the experimental values of $M_H$ and theoretical calculations in Eq.~\eqref{MH-ren-sub-OPE} for $H=B,\,D$.
In this fit,  a more accurate result for 
$[m_h]_{\rm PV}$ allows us to determine the nonperturbative matrix elements more accurately
(unless the neglected higher power corrections become significant).
To this end, we aim at accurate calculation of $[m_h]_{\rm PV}$ in Sec.~\ref{sec3}.

We would like to emphasize that Eqs.~\eqref{MH-ren-sub-OPE} and~\eqref{hyp-fine} 
are the same form between $H=B,\,D$ ($h=b,\,c$),
and, in particular, the non-perturbative matrix elements are identical.
This is because 
the matrix elements are defined to be independent of the mass of the heavy quark (infinite mass limit).
In order to subtract renormalons of $m_b$ and $m_c$ properly,
the same light quark theory should be used.
We choose $n_f=3$ theory,
i.e., we assume that up, down and strange quarks are massless 
and charm and bottom quarks are massive ($m_c,\,m_b\gg\LMS$). 
\footnote{It is known that the $n_f=3$ theory can describe the renormalon structure more properly than the $n_f=4$ theory. 
Physically, internal massive quarks in loops do not contaminate the IR structure. 
That is, it can be understood that internal charm quarks 
with non-zero (finite) masses do not contribute to the renormalon divergence of the bottom quark pole mass.
}

\subsection{\boldmath $u=1$ renormalon of $\bar{C}_{\bar{Q}Q}^\Gamma$
and its subtraction
}
\label{sec2.3}
In this subsection 
we discuss the detailed form of 
the $u=1$ renormalon uncertainty of $\bar{C}_{\bar{Q}Q}^\Gamma$
and consider an efficient way to subtract it.
Cancellation of the $u=1$ renormalon in the PV prescription requires the condition
\be
0={\rm Im}\bigg[\big[\bar{C}_{\bar{Q}Q}^\Gamma\big]_\pm\Big(1-\frac{\mu_\pi^2}{2m_b^2}\Big)\bigg]_{{\cal O}(\LMS^2/m_b^2)}
=\pm\,{\rm Im}\,\big[\bar{C}_{\bar{Q}Q}^\Gamma\big]_{{\cal O}(\LMS ^2/m_b^2)}-\big[\bar{C}_{\bar{Q}Q}^\Gamma\big]_{\rm PV}\frac{{\rm Im}\,\mu_\pi^2}{2\,\big[m_b\big]_{\rm PV}^2},
\ee
leading to
\be
{\rm Im}\,\big[\bar{C}_{\bar{Q}Q}^\Gamma\big]_{{\cal O}(\LMS^2/m_b^2)}\propto\big[\bar{C}_{\bar{Q}Q}^\Gamma\big]_{\rm PV}\frac{\LMS^2}{\big[m_b\big]_{\rm PV}^2},
\label{ImCbar}
\ee
where we used ${\rm Im}\,\mu_\pi^2\propto\LMS^2$. 
Eq.~\eqref{ImCbar} shows that the $u=1$ renormalon of $\bar{C}_{\bar{Q}Q}^\Gamma$ 
deviates from the integer power of $\LMS/m_b$,
because $\big[\bar{C}_{\bar{Q}Q}^\Gamma\big]_{\rm PV}$ 
is a function of $\alpha_s(m_b)$, which gives rise to complicated dependence 
on $\LMS/m_b$.
We note that, in the DSRS method, when 
a renormalon uncertainty has an integer power behavior, i.e.,
$(\LQ/Q)^n$ (where $n$ is an integer and $Q$ a typical scale of an observable),
we can use a simple formula.\footnote{
An extended formula is available to subtract the renormalon uncertainty of the form $f(\alpha_s(Q^2)) (\LQ/Q)^n$, where $f(\alpha_s(Q^2))$ is given as a series in $\alpha_s(Q^2)$.
In this paper, however, we exclusively use the simple formula for simplicity of analysis.
When there remains unremoved renormalon uncertainties due to the use of the simple formula,
effects of the residual renormalon are estimated and taken into account in our error budget. 
}
We then introduce a function $F=\log\big[\bar{C}_{\bar{Q}Q}^\Gamma\big]$
instead of $\bar{C}_{\bar{Q}Q}^\Gamma$ itself.
It turns out that the $u=1$ renormalon of $F$ has an integer power as follows.
\bea
{\rm Im}\,F&=&{\rm Im}\,\log\big[\bar{C}_{\bar{Q}Q}^\Gamma\big]
={\rm Im}\,\log\bigg[1\pm i\frac{{\rm Im}\,\big[\bar{C}_{\bar{Q}Q}^\Gamma\big]_{{\cal O}(\LMS^2/m_b^2)}}{\big[\bar{C}_{\bar{Q}Q}^\Gamma\big]_{\rm PV}}+{\cal O}(\LMS^3/m_b^3)\bigg]\non
&\propto&\frac{\LMS^2}{\big[m_b\big]_{\rm PV}^2}\big(1+{\cal O}(\LMS/m_b)\big).
\eea
Based on this observation, we consider $F$ and apply the DSRS method to separate $u=1$ renormalon from it.\footnote{
Although we reduced the renormalon uncertainty to the form $(\LMS/\big[m_b\big]_{\rm PV})^2$,
the complete subtraction of the $u=1$ renormalon is still not possible by the simple formula of the DSRS method
for the following reason. 
Complete subtraction of a renormalon is possible 
when it behaves as $(\LMS/Q)^n$, where $Q$ is {\it{the typical scale we choose in the DSRS calculation}}.
Then it is reasonable to choose $Q=\big[m_b\big]_{\rm PV}$, but this is practically difficult; 
when we do so we need to give perturbative series in terms of $\big[m_b\big]_{\rm PV}$,
which is conceptually different from $m_b$. The way to do this is not clear for us at present.
We then choose $Q=\mbar_b$, in which case the renormalon form is not
an integer power 
of $1/\mbar_b$.
We estimate the uncertainty from this incompleteness in Sec.~\ref{sec4.1}.
}
Then $\big[\bar{C}_{\bar{Q}Q}^\Gamma\big]_{\rm PV}$ can be reproduced by the exponentiation of $F$.

\section{Determination of HQET parameters by renormalon subtraction}
\label{sec3}
\subsection{PV masses in the DSRS method}
\label{sec3.1}
We determine the PV masses 
from the $\msbar$ masses 
using the DSRS method.
The perturbative relation between the pole mass $m_h$ and the $\msbar$ mass $\mbar_h$ of the heavy quark is given by 
\be
\frac{m_h-\mbar_h}{\mbar_h}\equiv\delta_h(\mbar_h)=\sum_{n=0}^\infty d_n^{(h)}(L_{\mbar})\alpha_s(\mu^2)^{n+1}.
\label{rel-pole-msbar}
\ee
Here, $L_{\mbar}=\log(\mu^2/\mbar_h^2)$, 
and $\mbar_h=m_h^{\msbar}(m_h^{\msbar})$.
$d_n^{(h)}(L_{\mbar})$ is a polynomial of $L_{\mbar}$ and can be determined by comparing the coefficient of $\alpha_s(\mu^2)^{n+1}$ on both sides of the relation
\be
\sum_{n=0}^\infty d_n^{(h)}(L_Q) \alpha_s(\mu^2)^{n+1}=e^{\hat{H}\log(\mu^2/\mbar_h^2)}\sum_{n=0}^\infty d_n^{(h)} \alpha_s(\mu^2)^{n+1},
\label{dn-Lmbar}
\ee
where $d_n^{(h)}=d_n^{(h)}(0)$.
The operator $\hat{H}$ is given with the QCD beta function $\beta$ by
\be
\hat{H}=-\beta(\alpha_s(\mu^2))\frac{\partial}{\partial\alpha_s(\mu^2)},~~~~\beta(\alpha_s)=\mu^2\frac{d\alpha_s(\mu^2)}{d\mu^2}=-\sum_{i=0}^\infty b_i\alpha_s(\mu^2)^{i+2},
\label{H}
\ee
which operates on the $\alpha_s(\mu^2)$ expansion on the right hand side of Eq.~\eqref{dn-Lmbar}.
In this paper, we use the five-loop beta function \cite{Baikov:2016tgj,Baikov:2017ayn,Herzog:2017ohr, Luthe:2017ttg} and the five-loop running coupling in practice.
The coefficients $d_n^{(h)}$ with massless light quarks are calculated up to $n=3$ 
\cite{Tarrach:1980up,Gray:1990yh,Chetyrkin:1999ys,Melnikov:2000qh,
Melnikov:2000zc,Marquard:2015qpa,Marquard:2016dcn}
and the contribution of massive internal quark,
i.e. the massive charm quark contribution to the mass relation of the bottom quark,
or, the 
massive bottom quark contribution to the mass relation of the charm quark,
is known up to $n=2$~\cite{Bekavac:2007tk,Fael:2020bgs}.
Although the massive internal quark contribution at $n=3$
is not available at present,
we expect that the $n=3$ coefficients can be well approximated by the known $n=3$ result in the massless case
based on the following observation.
First we consider the bottom quark mass relation
with the massive internal charm quark up to $n=2$:
\begin{align}
\delta_b(\mbar_b)
= 0.4244\,\alpha_s +  1.037\, \alpha_s^2 + 3.744\,\alpha_s^3
+{\cal O}(\alpha_s^4),
\label{mb-3f-tmp}
\end{align}
where 
$\alpha_s=\alpha_s^{(3)}(\mbar_b^2)$ with $\mbar_b=\mbar_b^{(5)}$.
The mass relation for the charm quark including (not-completely-decoupled) 
effects of the bottom quark is given by
\be
\delta_c(\mbar_c)
= 0.4244\, \alpha_s +1.044\,\alpha_s^2 +  3.757 \,\alpha_s^3
+{\cal O}(\alpha_s^4),
\label{mc-3f-tmp}
\ee
where $\alpha_s=\alpha_s^{(3)}(\mbar_c^2)$ with $\mbar_c=\mbar_c^{(4)}$.
On the other hand,
the mass relation with only three massless internal quarks (without massive quarks),
known up to $n=3$,
is given by
\be
\delta_h(\mbar_h)|_{\rm massless}
=0.4244\,\alpha_s+ 1.046\,\alpha_s^2+3.751\,\alpha_s^3+17.44\,\alpha_s^4+{\cal O}(\alpha_s^5),
\label{eq:mh0}
\ee
where $\alpha_s=\alpha_s^{(3)}(\mbar_h^2)$.
Comparing Eqs.~\eqref{mb-3f-tmp}, \eqref{mc-3f-tmp}, \eqref{eq:mh0},
they are very close;
the massive internal quark effects
are below 1\% at NLO,
and below 0.2\% at NNLO if we use the 3-flavor coupling constant.
This behavior is actually expected  
theoretically \cite{Ball:1995ni,Ayala:2014yxa,Hayashi:2021vdq}.
Therefore, we expect that the N$^3$LO correction 
in Eqs.~\eqref{mb-3f-tmp} and \eqref{mc-3f-tmp} 
can be well approximated by
that of Eq.~\eqref{eq:mh0}
and use
\begin{align}
\delta_b(\mbar_b)
&= 0.4244\,\alpha_s +  1.037\, \alpha_s^2 + 3.744\,\alpha_s^3
+17.44\,\alpha_s^4+{\cal O}(\alpha_s^5)\,,
\label{mb-3f}
\\
\delta_c(\mbar_c)
&= 0.4244\, \alpha_s +1.044\,\alpha_s^2 +  3.757 \,\alpha_s^3
+17.44\,\alpha_s^4+{\cal O}(\alpha_s^5)
\,,
\label{mc-3f}
\end{align}
in our analysis. (See Ref.~\cite{Hoang:2017btd} as a preceding estimate for the fourth order coefficient.)
The perturbative coefficients 
exhibit the factorial growth due to IR renormalons.
We assume that 
$\delta_h$ contains the renormalons at $u=1/2,\,1,\cdots$, 
whose effect appears as
the imaginary part of ${\cal O}((\LMS^2/\mbar_h^2)^u)$.

Using the DSRS method,
we consider the dual transform to suppress the IR renormalons at $u=1/2,\,1,\,\cdots$ 
and give the dual space series $\tilde{\delta}_h$. 
Following the notation in Ref.~\cite{Hayashi:2023fgl}
and choosing the parameters as $(a, u')=(2, -1/2)$,
the dual transform of $\delta_h$ is given by
\be
\tilde{\delta}_h(p)
=\int_{t_0-i\infty}^{t_0+i\infty}\frac{dt}{2\pi i}\,e^{tp^2}t^{-1}{\delta}_h(\mbar_h=1/t)
=\sum_{n=0}^\infty \tilde{d}_n^{(h)}
(L_{p})\alpha_s(\mu^2)^{n+1},
\label{til-delta}
\ee
where $L_p=\log(\mu^2/p^4)$, and $\tilde{d}_n$'s can be read from
\be
\sum_{n=0}^\infty \tilde{d}_n^{(h)}
\alpha_s(\mu^2)^{n+1}
=\frac{1}{\Gamma(1-2\hat{H})}\sum_{n=0}^\infty {d}_n^{(h)}
\alpha_s(\mu^2)^{n+1}.
\ee
Renormalon suppression in the dual space can be understood by the dual transform of the (approximate) imaginary part of $\delta_h$,
\be
\int_{t_0-i\infty}^{t_0+i\infty}\frac{dt}{2\pi i}\,e^{tp^2}
\LMS^{2u}t^{2u-1}=\frac{(\LMS^2/p^4)^u}{\Gamma(1-2u)}, \label{mass-renormalon-suppression}
\ee
which is zero for $u=1/2,\,1,\,\cdots.$\footnote{
As noted previously, it has not been made clear whether the $u=1$ renormalon exists or not.
However, we note that even in the case where the $u=1$ renormalon is absent,
our analysis to suppress the $u=1$ renormalon is valid;
in Ref.~\cite{Hayashi:2021ahf} it has been demonstrated that
with a choice of the parameters $(a, u')$ to suppress non-existing renormalons
the DSRS calculation still converges to a correct PV result.
}
The explicit form of the dual-space series $\tilde{\delta}_h$ is given by
\be
\tilde{\delta}_b(p )= e^{\hat{H}\log(\mu^2/p^4)}
\bigg[0.4244\,\alpha_s+0.6865\,\alpha_s^2+0.6872\,\alpha_s^3 - 2.656\,\alpha_s^4+{\cal O}(\alpha_s^5)
\bigg],
\label{til-del-b}
\ee
for the bottom quark and 
\be
\tilde{\delta}_c(p)= e^{\hat{H}\log(\mu^2/p^4)}
\bigg[0.4244\,\alpha_s+0.6928\,\alpha_s^2+0.6906\,\alpha_s^3 -2.747\,\alpha_s^4+{\cal O}(\alpha_s^5)
\bigg],
\label{til-del-c}
\ee
for the charm quark.
Here $\alfs=\alfs^{(3)}(\mu^2)$ and we use the N$^4$LO beta function 
and inputs are set to the central values of 
\be
\mbar_b=4.18^{+0.03}_{-0.02}~{\rm GeV},~
\mbar_c=1.27\pm0.02~{\rm GeV},~
\alpha_s^{(5)}(M_Z^2)=0.1179\pm0.0009,
\label{PDG-inputs}
\ee
from PDG~\cite{ParticleDataGroup:2022pth}.
Throughout this paper
we use these values unless stated explicitly.
The convergence of the series is 
improved in the dual space (Eqs.~\eqref{til-del-b} and \eqref{til-del-c}) compared with the original ones (Eqs.~\eqref{mb-3f} and \eqref{mc-3f}).
%
Using the formula in Ref.~\cite{Hayashi:2023fgl}, the PV mass $\big[m_h\big]_{\rm PV}$ is calculated as
\be
\big[m_h\big]_{\rm PV}=\mbar_h\Big(1+\big[\delta_h(\mbar_h)\big]_{\rm PV}\Big),
\label{mhPV-LB-DSRS}
\ee
where
\bea
\big[\delta(\mbar_h)\big]_{\rm PV}
&=&t\int_{0,\rm PV}^\infty dp^2\,e^{-tp^2}\tilde{\delta}_h(p)\non
&=&\frac{1}{\mbar_h}\int_{0,\rm PV}^\infty dp^2\,e^{-p^2/\mbar_h}
\sum_{n=0}^\infty \tilde{d}^{(h)}_n(L_p)\alpha_s(\mu^2)^{n+1}.
\label{del-PV}
\eea
$\big[\delta(\mbar_h)\big]_{\rm PV}$ can be obtained with good accuracy as more perturbative coefficients are used.
This is because, by setting $\mu\propto p^2$, we can take the integration contour such that $\alpha_s$ is always small and hence 
the integrand in Eq.~\eqref{del-PV} is a convergent series.

We note that the dual transform eliminates the $u=1$ renormalon of the form 
$\mbar_h\times\LMS^2/\mbar_h^2$, which is different from the actual $u=1$ renormalon in Eq.~\eqref{mhpm}.
We estimate the uncertainty caused by the residual renormalon uncertainty.
The imaginary part by the $u=1$ renormalon in Eq.~\eqref{mhpm} can be expressed as
\be
\frac{\LMS^2}{\big[m_h\big]_{\rm PV}}=\frac{\LMS^2}{\mbar_h}\frac{\mbar_h}{\big[m_h\big]_{\rm PV}}=\frac{\LMS^2}{\mbar_h}\bigg(1-d_0^{(h)}\alpha_s(\mbar_h^2)+{\cal O}\Big(\alpha_s(\mbar_h^2)^2\Big)\bigg),
\label{sub-ren}
\ee
where $\big[m_h\big]_{\rm PV}$ is expanded in $\alpha_s(\mbar_h^2)$ using the pole-$\msbar$ mass relation.\footnote{
While $\big[m_h\big]_{\rm PV}$ is a well-defined quantity where renormalons are subtracted, 
its perturbative expansion again has a factorially divergent behavior.
We expect that if the perturbative expansion is truncated at the first few orders as in Eq.~\eqref{sub-ren}
it gives a reasonable estimate, not affected strongly by renormalons.
}
The dual transform defined by Eq.~\eqref{til-delta} eliminates the contribution of the form 
\be
{\rm (const.)}\times\mbar_h\bigg(\frac{\LMS}{\mbar_h}\bigg)^{2u},
\ee
for $u=1/2,\,1,\,\cdots$,
and thus the ${\cal O}(\alpha_s(\mbar_h^2)\LMS^2/\mbar_h)$ uncertainty remains unremoved.
The typical size of the uncertainty for $\big[m_h\big]_{\rm PV}$ given by our calculation procedure
is estimated as
\be
\frac{\LMS^2}{\mbar_b}d_0^{(b)}\alpha_s(\mbar_b^2)\approx\frac{0.3^2~{\rm GeV}^2}{4.18~{\rm GeV}}\times\frac{4}{3\pi}\times0.21\approx{\cal O}(2~{\rm MeV}),
\label{sub-ren-mb}
\ee
for the bottom quark
and
\be
\frac{\LMS^2}{\mbar_c}d_0^{(c)}\alpha_s(\mbar_c^2)\approx\frac{0.3^2~{\rm GeV}^2}{1.27~{\rm GeV}}\times\frac{4}{3\pi}\times0.39 \approx{\cal O}(10~{\rm MeV}),
\label{sub-ren-mc}
\ee
for the charm quark.
These uncertainties are small
compared to the total uncertainties in our determination as we see below.
In addition, we expect that the normalization constant of the $u=1$ renormalon is small \cite{Neubert:1996zy,Ayala:2019hkn}
and the above estimates assuming the normalization constant to be $\mathcal{O}(1)$
would be conservative.

For later use, 
we estimate the N$^4$LO correction to $\delta_h$ and to its dual transform $\tilde{\delta}_h$.
The procedure for this estimate is briefly explained in App.~\ref{App.B}.
We use
\be
d_4^{(b,c)\rm est}\approx111.3\pm7.72, \label{d4est}
\ee
and 
\be
\tilde{d}_4^{(b)\rm est}\approx-15.25\pm7.72 , 
\quad{}
\tilde{d}_4^{(c)\rm est}\approx-15.54\pm7.72
\,,
\ee
for the dual-space.
The uncertainty is from the determination of the normalization constant of the $u=1/2$ renormalon.

\begin{figure}[tbp]
\centering
\includegraphics[width=14cm]{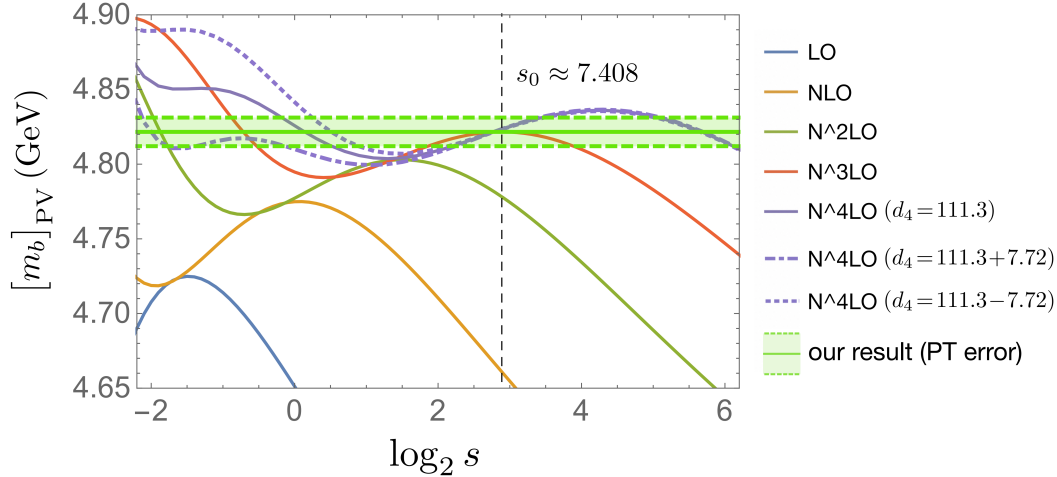}
\caption{\small
Determination of $\big[m_b\big]_{\rm PV}$ using the DSRS method.
The colored curves represent the scale dependence of N$^k$LO results for $k=0,1,2,3,4$.
The input parameters are $\mbar_b=4.18~{\rm GeV}$, $\mbar_c=1.27~{\rm GeV}$ and $\alfs(M_Z^2)=0.1179$.
The green line and shaded area show the determined value using the N$^3$LO result.
}
\label{mbPV-msbar}
\end{figure}

Now we evaluate the PV mass of the bottom quark.
Fig.~\ref{mbPV-msbar} shows the scale dependence of $\big[m_b\big]_{\rm PV}$ when we set $\mu=sp^2$ 
($L_p=\log{s^2}$) and truncate the series in Eq.~\eqref{del-PV} at $\alpha_s^{k+1}$
for $h=b$; see Eq.~\eqref{til-del-b}. The colored curves represent the results for $k=0,\,1,\,2,\,3,\,4$ when the inputs are set to the central value of the PDG values.
As a reference, the estimated result for $k=4$ by using $d_4^{(b)\rm est}$ is given by the purple curve and the uncertainty is displayed by the dotted and dot-dashed curves.
Since the N$^3$LO curve is stable at the stationary point around $\log_2s=3$, 
we give the central value by the green line and the uncertainty by the band as in Fig~\ref{mbPV-msbar}.
The uncertainty is estimated by the scale variation from $s_0/2$ to $2s_0$ 
around the stationary point $s=s_0\approx7.408$.\footnote{
Although $s_0 \approx 7.408$ is relatively large,
one can confirm that the error band overlaps with the values around $s=1$ (or $\log_2{s}=0$).
}
Since there are multiple stationary points,
we choose the one about which the PV mass is most stable against 
the scale variation by a factor 2 or 1/2.
We obtain 
\bea
\big[m_b\big]_{\rm PV}
&=&4.822\,(10)_{\rm PT}\,(2)_{{\rm sub}\,u=1}\,(33)_{\mbar_b}\,(0)_{\mbar_c}\,(8)_{\alpha_s}~{\rm GeV}\non
&=&4.822\,(10)_{\rm th}\,(34)_{\rm input}~{\rm GeV}=4.822\,(36)~{\rm GeV},
\label{mbPVres}
\eea
where the central value is given by the value of the N$^3$LO curve at $s=s_0\approx7.408$.
The number in the first brackets in the first line denotes the uncertainty from the scale dependence discussed above.
The second one is estimated by the subleading effect of the $u=1$ renormalon considered in Eq.~\eqref{sub-ren-mb}.
The third, fourth and fifth ones represent the uncertainties from 
the input PDG values of $\mbar_b$, $\mbar_c$ and $\alpha_s(M_Z^2)$, respectively.
In the second line, the uncertainties are combined in quadrature.
The first two uncertainties are accounted for the theoretical uncertainty
and the others are for the input uncertainty.
It can be seen that the theoretical uncertainty is small enough ($\sim 0.2 \%$)
compared to the case where the $u=1/2$ renormalon remains;
in this case an $\mathcal{O}(\LMS/\mbar_b) (\sim 7~\%)$  uncertainty is expected.
This shows a successful subtraction of the renormalon by the DSRS method.
One can also see that the theoretical uncertainty is smaller than the input uncertainty.


Let us make a brief comment on
the input parameter $\mbar_b$.
The result in Eq.~\eqref{mbPVres}, in terms of the PDG value, is a conservative estimate.
The Flavor Lattice Averaging Group (FLAG)~\cite{FlavourLatticeAveragingGroupFLAG:2021npn} 
has recently reported
\be
\mbar_b=4.171(20)~{\rm GeV}\quad {\rm from}\quad N_f=(2+1)\,{\rm lattice},
\label{FLAG2021mb0}
\ee
and
\be
\mbar_b=4.203(11)~{\rm GeV}\quad {\rm from}\quad N_f=(2+1+1)\,{\rm lattice},
\label{FLAG2021mb}
\ee
where $N_f$ denotes the number of active quarks in the lattice simulations.
By adopting these values, we obtain 
\be
\big[m_b\big]_{\rm PV}=4.812\,(22)_{\mbar_b}~{\rm GeV}\quad({\rm N}^3{\rm LO},\,N_f=(2+1)),
\ee
and
\be
\big[m_b\big]_{\rm PV}=4.847\,(12)_{\mbar_b}~{\rm GeV}\quad({\rm N}^3{\rm LO},\,N_f=(2+1+1)),
\ee
respectively.
The uncertainties from other sources, as in Eq.~\eqref{mbPVres}, are almost unchanged.
Note that there is a slight
inconsistency between the values of $\mbar_b$ 
by the lattice simulation for $N_f=2+1$ and $N_f=2+1+1$. 
The values given here are only for comparison
and not used in our actual analysis.

\begin{figure}[tbp]
\centering
\includegraphics[width=15cm]{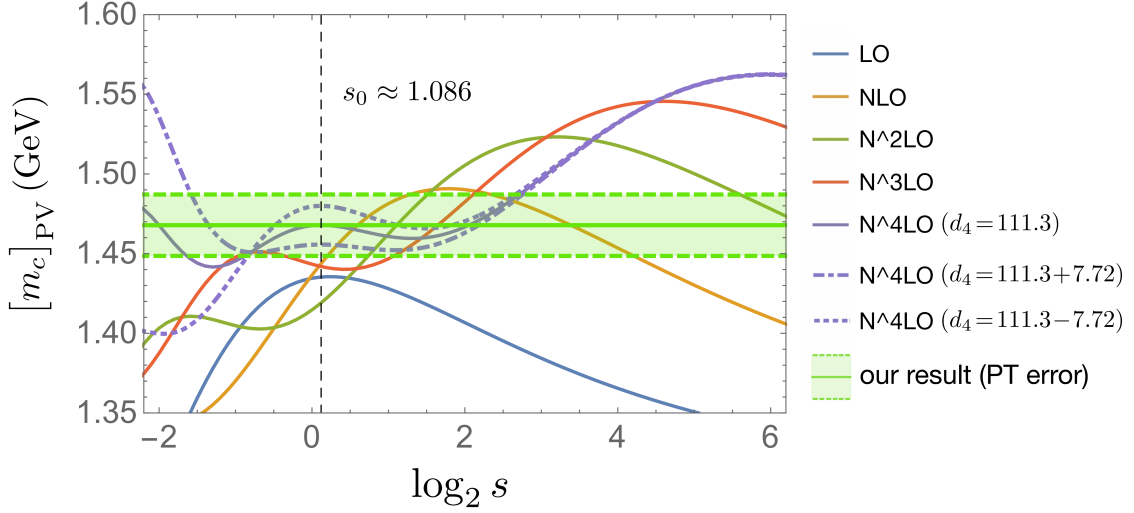}
\caption{\small
Determination of $\big[m_c\big]_{\rm PV}$ from the actual perturbative expansion using the DSRS method.
The colored curves represent the scale dependence of N$^k$LO results for $k=0,1,2,3,4$.
Inputs for calculations are the same as Fig.~\ref{mbPV-msbar}.
The green line and shaded area show the determined value using the estimated N$^4$LO result.
}
\label{mcPV-msbar}
\end{figure}

Next we evaluate the PV mass of the charm quark.
The procedure is similar to that of the bottom quark,
but unlike the bottom quark case, we find that at present, the number of known perturbative coefficients
 is insufficient to determine $\big[m_c\big]_{\rm PV}$ precisely.
In this analysis, we therefore use $d_4^{(c)\rm est}$ to determine $\big[m_c\big]_{\rm PV}$.
Fig.~\ref{mcPV-msbar} shows the scale dependence of $\big[m_c\big]_{\rm PV}$ 
when we set $\mu=sp^2$ and truncate $\tilde{\delta}_c$ at $\alpha_s^{k+1}$.
The uncertainty from $d_4^{(c)\rm est}$ is shown as dotted and dot-dashed curves.
We obtain 
\bea
\big[m_c\big]_{\rm PV}
&=&1.468\,(19)_{\rm PT}\,(12)_{d_4}\,(10)_{{\rm sub}\,u=1}\,(0)_{\mbar_b}\,(24)_{\mbar_c}\,(4)_{\alpha_s}~{\rm GeV}\non
&=&1.468\,(25)_{\rm th}\,(25)_{\rm input}~{\rm GeV}=1.468\,(35)~{\rm GeV},
\label{mcPVres}
\eea
where the central value is given by the value of the N$^4$LO solid curve at $s=s_0\approx1.086$.
The second uncertainty comes from the uncertainty in the estimate of $d_4^{(c)\rm est}$,
and the rest is the same as the bottom quark case.
The theoretical uncertainty (composed of the first three uncertainties)
is comparable to the input uncertainty.
As in the case of the bottom quark,
we present the outputs using FLAG~\cite{FlavourLatticeAveragingGroupFLAG:2021npn} values:
\begin{align}
&\mbar_c=1.275(5)~{\rm GeV}\quad {\rm from}\quad N_f=(2+1)\,{\rm lattice},
\\
&\mbar_c=1.278(13)~{\rm GeV}\quad {\rm from}\quad N_f=(2+1+1)\,{\rm lattice},
\label{FLAG2021mc}
\end{align}
which give 
\begin{align}
&\big[m_c\big]_{\rm PV}=1.474\,(6)_{\mbar_c}~{\rm GeV}\quad({\rm N}^4{\rm LO},\,N_f=(2+1)),
\\
&\big[m_c\big]_{\rm PV}=1.478\,(16)_{\mbar_c}~{\rm GeV}\quad({\rm N}^4{\rm LO},\,N_f=(2+1+1)),
\end{align}
respectively.

\subsection{\boldmath Determination of $\bar{\Lambda}$ and $\mu_\pi^2$}
\label{sec3.2}
In this subsection
we determine 
the HQET parameters in Eq.~\eqref{MH-ren-sub-OPE}.
We use the PDG values of the meson masses \cite{ParticleDataGroup:2020ssz} given by
\be
M_B=\frac{5.27965 + 5.27934}{2}~{\rm GeV},\,M_{B^*}=5.32470~{\rm GeV}.
\ee
and
\be
M_D=\frac{1.86484 + 1.86966}{2}~{\rm GeV},\,M_{D^*}=2.00685~{\rm GeV},
\ee
where the uncertainties assigned by PDG are negligible in our analysis.
Using the results of Eqs.~\eqref{OPE-MH}, \eqref{MH-ren-sub-OPE}, \eqref{mbPVres} 
and \eqref{mcPVres}, we obtain
\bea
\big[\bar{\Lambda}\big]_{\rm PV}
&=&0.486\,(16)_{\rm PT}\,(5)_{d_4}\,(5)_{{\rm sub}\,u=1}\,(6)_{1/m_h^2}\,(48)_{\mbar_b}\,(11)_{\mbar_c}\,(13)_{\alpha_s}~{\rm GeV}\non
&=&0.486\,(19)_{\rm th}\,(50)_{\rm input}~{\rm GeV}=0.486\,(54)~{\rm GeV},
\label{res-Lbar}
\eea
and
\bea
\big[\mu_\pi^2\big]_{\rm PV}
&=&0.05\,(9)_{\rm PT}\,(5)_{d_4}\,(4)_{{\rm sub}\,u=1}\,(5)_{1/m_h^2}\,(14)_{\mbar_b}\,(11)_{\mbar_c}\,(5)_{\alpha_s}~{\rm GeV}^2\non
&=&0.05\,(12)_{\rm th}\,(18)_{\rm input}~{\rm GeV}^2=0.05\,(22)~{\rm GeV}^2.
\label{res-mupi2}
\eea
The `PT' uncertainty is estimated as follows.
Denoting $\big[m_b \big]_{\rm PV}=4.822+ 0.010 \cdot \delta^{(b)}_{\rm PT}$~GeV
and $\big[m_c \big]_{\rm PV}=1.468+ 0.019 \cdot  \delta^{(c)}_{\rm PT}$~GeV (see Eqs.~\eqref{mbPVres} and \eqref{mcPVres}),
we set only one $\delta_{\rm PT}$ to either $+1$ or $-1$, 
(whereas the other $\delta$'s are set to zero) 
and see the difference of the determined non-perturbative parameters from the central value.
We take the maximum difference  as our `PT' uncertainty among the four cases.
The uncertainties regarding $d_4$ and `sub $u=1$'  are estimated in a similar manner. 
(The $d_4$ uncertainty is relevant only to $\big[m_c \big]_{\rm PV}$.)
In estimating the fourth uncertainty related to the truncation of the $1/m_h$ expansion, 
we add a term $\LMS^3/\mbar_h^2$ with its coefficient either $-1, 0 , +1$
and take the maximum difference among the cases we turn on only one 
of the coefficients.
We estimate the input uncertainties similarly but noting the point that 
the variation of $\big[m_b \big]_{\rm PV}$ and $\big[m_c \big]_{\rm PV}$
are not independent in this case; for instance a larger $\alpha_s(M_Z^2)$ gives a lager $\big[m_b \big]_{\rm PV}$
and a smaller $\big[m_c \big]_{\rm PV}$.

The theoretical uncertainty of $\bar{\Lambda}$
is small compared to ${\cal O}(\LMS)\sim300~{\rm MeV}$, reflecting the subtraction of the $u=1/2$ renormalon.
For $\mu_\pi^2$, however, even the theoretical uncertainty has a magnitude of ${\cal O}(\LMS^2)$, as if the $u=1$ renormalon remained in our calculation.
We believe that the small number of the known perturbative coefficients results in a large perturbative uncertainty of $\big[m_c\big]_{\rm PV}$, which is reflected in $\mu_\pi^2$.
The perturbative uncertainty will be reduced 
once higher-order perturbative corrections are incorporated provided the renormalons are properly subtracted, 
and the theoretical uncertainty should eventually be reduced 
to an accuracy of  $\sim0.04~{\rm GeV}^2$, the scale of the residual $u=1$ renormalon, cf., Eq.~\eqref{sub-ren-mc}.
It is also important to check the effects of higher power corrections,
after subtracting higher renormalons beyond $u=1$.
This will be the task of our future study.

Let us briefly compare
our results
with other studies.
In Ref.~\cite{FermilabLattice:2018est},
\bea
\big[\bar{\Lambda}\big]_{\rm PV}=0.435(31)~{\rm GeV}\,,
~~~
\big[\mu_\pi^2\big]_{\rm PV}=0.05(22)~{\rm GeV}^2\,
\eea
are reported.
In this fit,
only the $u=1/2$ renormalon is subtracted, while
the ${\cal O}(1/m_h)$ corrections are included.
They argue that the uncertainty of $\mu_\pi^2$ originates from the
$u=1$ renormalon.
We have, however, a different interpretation.
In our analysis,
the $u=1$ renormalon is
already subtracted but the result still has the same order uncertainty as theirs.
Hence, it would be simply due to insufficiency of perturbative order
at the current status.
The subtraction of the $u=1$ renormalon would matter only
when higher order perturbative coefficients are available.
In Ref.~\cite{Ayala:2019hkn} 
\bea
\big[\bar{\Lambda}\big]_{\rm PV}=
477(\mu)^{-8}_{+17}(Z_m)^{+11}_{-12}(\alfs)^{-8}_{+9}({\cal O}(1/m_h))^{+46}_{-46}
~{\rm MeV}\,,
\eea
is obtained,
in which only the $u=1/2$ renormalon is subtracted and
the ${\cal O}(1/m_h)$ corrections are neglected (including the $\mu_\pi^2$ term).
The above two results 
are consistent with our determination
within the assigned uncertainties.

\section{\boldmath Determination of $|V_{cb}|$}
\label{sec4}
\subsection{\boldmath Subtracting $u=1$ renormalon from the decay width OPE}
\label{sec4.1}
The OPE of the inclusive semileptonic $B$ decay width is
given by Eq.~\eqref{OPE-Gamma-mbar},
and we apply the DSRS method to 
the most important part in the OPE,
that is
$\bar{C}_{\bar{Q}Q}^\Gamma$.
It is given by
\bea
\bar{C}_{\bar{Q}Q}^\Gamma
&=&\Big(\frac{m_b}{\mbar_b}\Big)^5C_{\bar{Q}Q}^\Gamma\big(m_b(\mbar_b),\rho={m_c(\mbar_c)}/{m_b(\mbar_b)}\big)\non
&=&(1+\delta_b(\mbar_b))^5\sum_{n=0}^\infty X_n\big(\rho=\bar{\rho}\,\frac{1+\delta_c(\mbar_c)}{1+\delta_c(\mbar_b)}\big)\alpha_s(m_b(\mbar_b)^2)^n\non
&=&\sum_{n=0}^\infty \bar{X}_n(\bar{\rho}) \alpha_s(\mbar_b^2)^n,
\label{barC-def}
\eea
where $\delta_c(\mbar_c)$ and $\alpha_s(m_b(\mbar_b)^2)$ are expanded in $\alpha_s(\mbar_b^2)$.
In rewriting, we change the number of flavors of the coupling constant to $n_f=3$.
The explicit form of $\bar{C}_{\bar{Q}Q}^\Gamma$ is given by
\be
\bar{C}_{\bar{Q}Q}^\Gamma\approx0.5114\big(1+1.593\,\alpha_s+ 3.579\,\alpha_s^2+8.894\,\alpha_s^3+{\cal O}(\alpha_s^4)\big),
\label{barC-conv}
\ee
where $\alpha_s=\alpha_s^{(3)}(\mbar_b^2)$.
Due to the fifth power of $m_b$ contained in $\Gamma$, the perturbative series shows a slow convergence even after cancellation of the $u=1/2$ renormalon. 

Let us take the logarithm of $\bar{C}_{\bar{Q}Q}^\Gamma$ as
\bea
F&=&\log\Big[\bar{C}_{\bar{Q}Q}^\Gamma\Big]
=\log\Big[\bar{X}_0+\sum_{n=1}^\infty \bar{X}_n\alpha_s(\mbar_b^2)^{n}\Big]
\label{logC}
\\
&\equiv&\log\bar{X}_0+\sum_{n=0}^\infty \chi_n\alpha_s(\mbar_b^2)^{n+1},
\label{F-def}
\eea
where we use $\log(1+x)=\sum_{n=1}^\infty(-1)^{n+1}x^n/n$ to construct $\chi_n$.
The explicit form of $F(\mbar_b)$ is 
\be
F\approx-0.6707+1.593\,\alpha_s+2.311\,\alpha_s^2+4.540\,\alpha_s^3+{\cal O}(\alpha_s^4),
\label{F-exp}
\ee
where $\alpha_s=\alpha_s^{(3)}(\mbar_b^2)$.
We find that 
the logarithmic transform partially cancels the large coefficients we have in Eq.~\eqref{barC-conv}.

Choosing the parameters of the DSRS method as
$(a,u')=(1,-1)$,
the dual transform of $F$ is defined by
\bea
\widetilde{F}(p)
&=&\int_{t_0-i\infty}^{t_0+i\infty}\frac{dt}{2\pi i}\,e^{tp^2}t^{-1}F(\mbar_b=1/\sqrt{t})\non
&=&\log\bar{X}_0+e^{\hat{H}\log(\mu^2/p^2)}\sum_{n=0}^\infty\tilde{\chi}_n\alpha_s(\mu^2)^{n+1},
\label{til-F-def}
\eea
in which the $u=1$ renormalon is suppressed
(Note that the mass dimension of the dual variable $p^2$
is different from that of the PV mass, cf.\ Eq.~\eqref{til-delta}.)
$\tilde{\chi}_n$ can be read from
\be
\sum_{n=0}^\infty\tilde{\chi}_n\alpha_s(\mu^2)^{n+1}=\frac{1}{\Gamma(1-\hat{H})}\sum_{n=0}^\infty{\chi}_n\alpha_s(\mu^2)^{n+1},
\ee
and $\widetilde{F}$ is given by
\be
\widetilde{F}\approx-0.6707+1.593\,\alpha_s+1.653\,\alpha_s^2+1.185\,\alpha_s^3+{\cal O}(\alpha_s^4)\,,
\label{til-F-exp}
\ee
where $\alpha_s=\alpha_s^{(3)}(p^2)$.
This shows a better convergence behavior than Eq.~\eqref{F-exp}.

The inverse dual transform gives the regularized quantity $\big[F\big]_{\pm}$ by
\bea
\big[F(\mbar_b)\big]_{\pm}
&\equiv&\big[F(\mbar_b)\big]_{\rm PV}\pm i\,{\rm Im}\,F(\mbar_b)\non
&=&\log\bar{X}_0+\int_{C_\mp}\frac{dp^2}{\mbar_b^2}\,e^{-p^2/\mbar_b^2}
e^{\hat{H}\log(\mu^2/p^2)}
\sum_{n=0}^\infty\tilde{\chi}_n\alpha_s(\mu^2)^{n+1},
\eea
where the integral is evaluated numerically.
$C_{\mp}$ is the contour avoiding the Landau singularity of $\alpha_s(s^2 p^2)$ 
in the integrand (when $\mu^2$ is set 
proportional to $p^2$) slightly below/above the positive real axis.
We solve the RG equation of $\alpha_s$ along $C_{\pm}$.

We note that we suppress renormalons of the form 
\be
{\rm (const.)}\times\bigg(\frac{\LMS}{\mbar_b}\bigg)^{2u},
\ee
for $u=1,\,2,\,3,\cdots$ in this calculation.
This indicates that the ${\cal O}(\alpha_s(\mbar_b^2)\LMS^2/\mbar_b^2)$ uncertainty remains.
In a parallel manner to Eq.~\eqref{sub-ren-mb} or \eqref{sub-ren-mb},
we estimate its possible uncertainty for $|V_{cb}|$ (see `sub $u=1$' in Eq.~\eqref{res-Vcb-PDG}).

\begin{figure}[t]
\centering
\includegraphics[width=16cm]{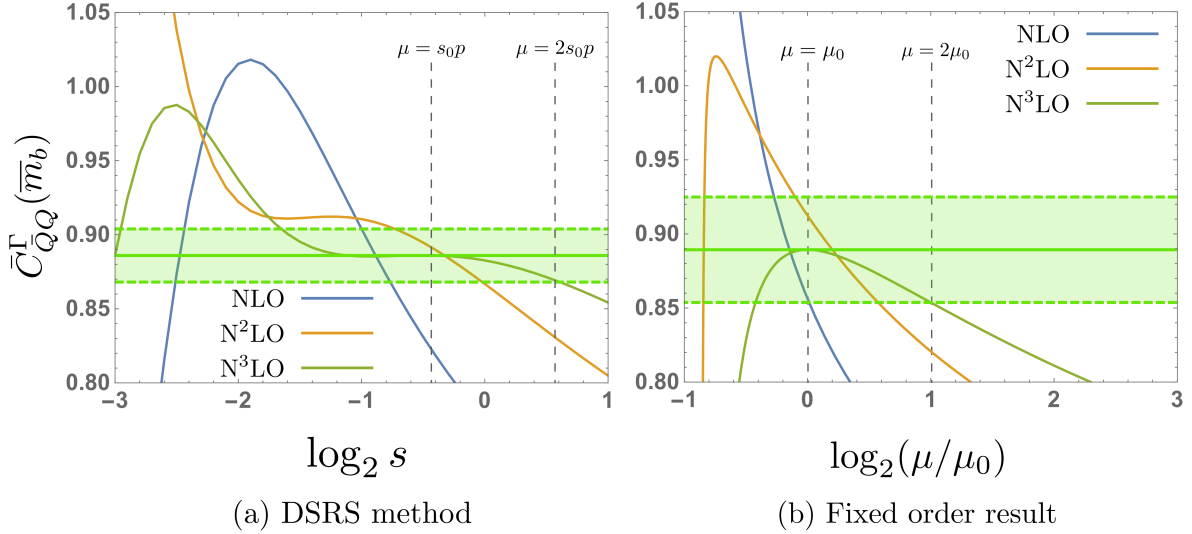}
\caption{\small
Comparing the scale dependence of $\bar{C}_{\bar{Q}Q}^\Gamma$ calculated by the DSRS method (left panel) 
and by the fixed-order perturbation (right panel).
The colored curves represent the scale dependence of N$^k$LO results for $k=1,2,3$.
The green lines and shaded areas exhibit the determined value using the N$^3$LO result 
and the uncertainty, respectively.
For the fixed order result, we do not choose $\mu=\mu_0/2$ to estimate the uncertainty 
because it gives an unreasonably large uncertainty because of the Landau pole singularity.
}
\label{Cbar-res}
\end{figure}

We compute
$\big[\bar{C}_{\bar{Q}Q}^\Gamma\big]_{\rm PV}$
and investigate the scale dependence.
Fig.~\ref{Cbar-res} shows the scale dependence of 
\be
\big[\bar{C}_{\bar{Q}Q}^\Gamma\big]_{\rm PV}=\bar{X}_0\,{\rm Re}\,\exp\bigg(\int_{C_\mp} \frac{dp^2}{\mbar_b^2}\,
e^{-p^2/\mbar_b^2}\sum_{n=0}^k \tilde{\chi}_n(\log s^2) \alpha_s(s^2p^2)^{n+1}\bigg),
\label{barC-PV-as3}
\ee
where $\tilde{\chi}(\log s^2)$ is determined by the relation similar 
to Eq.~\eqref{dn-Lmbar}.
The colored curves represent the results for $k=0,\,1,\,2$.
Around $s={\cal O}(1)$, we can see the flat region in the N$^3$LO result.
In this analysis, we find the minimal sensitivity scale $s=s_0$ such that the difference between $\big[\bar{C}_{\bar{Q}Q}^\Gamma\big]_{\rm PV}$ at $s=2s_0$ and $s=s_0/2$ is minimized.
For the N$^3$LO curve, we obtain $s_0\approx0.7624$.
At $s=s_0$, $\big[\bar{C}_{\bar{Q}Q}^\Gamma(\mbar_b)\big]_{\rm PV}$ is given by
\bea
\big[\bar{C}_{\bar{Q}Q}^\Gamma(\mbar_b)\big]_{\rm PV}
&=&0.511~{\rm Re}\,e^{(0.470 + 0.0507 i)_{{\cal O}(\alfs)} + (0.0864 + 0.0314 i) _{{\cal O}(\alfs^2)}+ (-0.00408 - 0.00337 i)_{{\cal O}(\alfs^3)}}
\non
\label{barC-PV-exp}
\\
&=&0.511|_{\rm LO}+0.306|_{\rm NLO}+0.0719|_{\rm N^2LO}-0.00338|_{\rm N^3LO},
\label{barC-PV-as}
\eea
where the N$^k$LO correction for $k=1,\,2,\,3$ in Eq.~\eqref{barC-PV-as} is calculated by the difference of the right hand side of Eq.~\eqref{barC-PV-exp} 
evaluated by truncating the exponent at ${\cal O}(\alpha_s^k)$ and at ${\cal O}(\alpha_s^{k-1})$.
We estimate the the perturbative uncertainty
by the difference between the values at $s=s_0$ and $s=2s_0$.
(The difference between the values at $s=s_0$ and $s=s_0/2$ is smaller.)
Then we obtain
\bea
\big[\bar{C}_{\bar{Q}Q}^\Gamma(\mbar_b)\big]_{\rm PV}
&=&0.886\pm0.016,
\label{barC-PV-err}
\eea
which is displayed by the green line and shaded area in Fig.~\ref{Cbar-res}.
Note that the size of the N$^3$LO correction in Eq.~\eqref{barC-PV-as} is, by accident, smaller
than the uncertainty in Eq.~\eqref{barC-PV-err}
which we adopt in our analysis.

Let us compare the DSRS result and the conventional perturbative expansion given by Eq.~\eqref{barC-def}.
The right panel of Fig.~\ref{Cbar-res} shows the scale dependence of the fixed-order perturbation result, which is given by
\be
\bar{C}_{\bar{Q}Q}^\Gamma(\mbar_b)\bigg|_{\rm fix}
=e^{\hat{H}\log(\mu^2/\mbar_b^2)}\sum_{n=0}^k \bar{X}_n\alpha_s(\mu^2)^{n+1},
\label{barC-RG-as3}
\ee
with $\mu=\mu_0$ and we truncate at ${\cal O}(\alpha_s^3)$ after expanding 
the exponential factor $e^{\hat{H}\log(\mu^2/\mbar_b^2)}$.
We can find a stationary point on the N$^3$LO
curve of the fixed-order result at $\mu=\mu_0\approx1.1387$~GeV.
The fixed order result at this scale is given by
\be
\bar{C}_{\bar{Q}Q}^\Gamma(\mbar_b)\bigg|_{\rm fix}=0.511|_{\rm LO} + 0.299|_{\rm NLO} + 0.0717|_{\rm N^2LO} + 0.00239|_{\rm N^3LO},
\ee
which appears to be almost the same as Eq.~\eqref{barC-PV-as}.
However, the value of $\bar{C}_{\bar{Q}Q}^\Gamma|_{\rm fix}$ changes 
rapidly when the scale is varied around the stationary point. 
This is because of the unphysical singularity of $\alpha_s$ at $\mu\sim\LMS$.
On the other hand, the DSRS result shows 
milder scale dependence around the stationary point than the fixed-order result,
which is partly
due to the removal of the unphysical singularity by the DSRS method.
The uncertainty estimated from the scale variation is about twice
larger than the DSRS result in Eq.~\eqref{barC-PV-err}.
Furthermore, it is observed that the N$^3$LO values at the stationary points in DSRS and the fixed order calculation
are close to each other.
We expect the good convergence of Eq.~\eqref{barC-PV-as3} to continue at higher orders 
as a consequence of the subtraction of the renormalons.

\subsection{\boldmath  $|V_{cb}|$ determination}
\label{sec4.2}
In this section, we determine $|V_{cb}|$ using the DSRS result of $\bar{C}_{\bar{Q}Q}^\Gamma$ in the previous section.
We use the renormalon-subtracted OPE of $\Gamma$ given by
\be
\Gamma=\frac{G_F^2|V_{cb}|^2}{192\pi^3}A_{ew}\overline{m}_b^5\bigg[\big[\bar{C}_{\bar{Q}Q}^\Gamma\big]_{\rm PV}\Big(1-\frac{\big[\mu_\pi^2\big]_{\rm PV}}{2\big[m_b\big]_{\rm PV}^2}\Big)+\bar{C}_{cm}^\Gamma\frac{\mu_G^2}{\big[m_b\big]_{\rm PV}^2}\bigg],
\ee
where we neglect the ${\cal O}({\LMS^3}/{m_b^3})$ OPE corrections and the 
residual $u=1$ renormalon effect of ${\cal O}(\alpha_s(\mbar_b)\LMS^2/m_b^2)$.
We use $\big[m_b\big]_{\rm PV}$, $\big[\bar{C}_{\bar{Q}Q}^\Gamma\big]_{\rm PV}$, $\bar{C}_{cm}^\Gamma$, $\big[\mu_\pi^2\big]_{\rm PV}$ and $\mu_G^2$ given by Eqs.~\eqref{mbPVres}, \eqref{barC-PV-err}, \eqref{CcmLO}, \eqref{res-mupi2}, and \eqref{res-muG2}, respectively.
The NLO correction to $\bar{C}_{cm}^\Gamma$ of ${\cal O}(\alpha_s)$ given by Eq.~\eqref{CcmNLO} is used to estimate the systematic uncertainty.
The experimental value of $\Gamma$ is obtained as follows.
In this analysis, we neglect the iso-spin breaking and assume that the semileptonic decay width of $B_0$
and that of $B^\pm$ is the same.
Then we have the semileptonic decay width $\Gamma$ by 
\be
\Gamma={\cal B}/\tau_B,
\ee 
where ${\cal B}={\cal B}(B\to X_c\ell \nu)$ represents the semileptonic branching ratio obtained for the admixture of $B^0/B^{\pm}$.
$\tau_B$ is the lifetime of the $B$ meson given by
\be
\tau_B=\frac{\tau_{B^{\pm}}+\tau_{B^0}}{2}+\frac{1}{2}(f^{+-}-f^{00}) (\tau_{B^{\pm}}-\tau_{B^0}),
\label{tauB-def}
\ee
as explained in Sec.~III~A of Ref.~\cite{Bauer:2004ve}.
Here $f^{+-}$ and $f^{00}$ are the fractions of the $B^+ B^-$ production and the $B^0 \bar{B}^0$ production from the $\Upsilon(4S)$ decay, respectively.
We give the value $\tau_B$ by
\bea
\tau_B
&=&\frac{(1.519\pm0.004)+(1.638\pm0.004)}{2}\times10^{-12}\, {\rm sec}\non
&=&(1.579\pm0.004)\times10^{-12}\, {\rm sec},
\label{tauB}
\eea
neglecting the second term in Eq.~\eqref{tauB-def}
since it is smaller than the uncertainty of $\tau_{B^{\pm}}$ or $\tau_{B^0}$.
See Ref.~\cite{HFLAV:2019otj} for the ratio $f^{+-}/f^{00}$.
In our analysis we use the value of ${\cal B}$,
\be
{\cal B}=(10.63\pm0.19)\%,
\label{Br}
\ee
which is given in Ref.~\cite{Bernlochner:2022ucr}  as the average of the Belle measurement \cite{Belle:2006kgy} and BaBar measurement \cite{BaBar:2009zpz}.\footnote{
In Ref.~\cite{Bernlochner:2022ucr}, results for $\cal B$ with a lepton energy cut are extrapolated to the full phase space by using
a theoretical calculation of the decay width in the kinetic mass scheme. 
Although our determination in the $\overline{\text{MS}}$ mass scheme is not completely independent of the kinetic mass scheme determination in this sense,
the scheme dependence of Eq.~\eqref{Br} is expected to be small enough because 
it is stated that the power corrections in the OPE are not significant in obtaining Eq.~\eqref{Br} \cite{Bernlochner:2022ucr}.
}
Our result of $|V_{cb}|$ in the $\msbar$ mass scheme is given by
\bea
|V_{cb}|
&=&0.0415\,(4)_{\rm PT}\,(_{-9}^{+6})_{\mbar_b}\,(4)_{\mbar_c}\,(2)_{\alpha_s}\,(1)_{\tau_B}\,(4)_{\cal B}\,(1)_{\mu_\pi^2}\,(0)_{\mu_G^2}\,(0)_{{\rm h.o.}\,C_{cm}}\,(1)_{1/m_b^3}\,(0)_{{\rm sub}\,u=1}\non
&=&0.0415\,(_{-12}^{+10}),
\label{res-Vcb-PDG}
\eea
where the brackets in the first line denote the systematic uncertainties.
The first uncertainty comes from the perturbative uncertainty estimated by Eq.~\eqref{barC-PV-err}.
The second, third and fourth uncertainties are caused by the uncertainties of the PDG inputs.
The fifth and sixth uncertainties come from the uncertainties of the experimental data in Eqs.~\eqref{tauB}~and~\eqref{Br}, respectively.
The last five uncertainties are related to the non-perturbative corrections, 
$\big[\mu_\pi^2\big]_{\rm PV}$ given by Eq.~\eqref{res-mupi2}, 
$\mu_G^2$ given by Eq.~\eqref{res-muG2},
the NLO correction to $\bar{C}_{cm}^\Gamma$,
the neglected ${\cal O}(\LMS^3/m_b^3)$ OPE correction
and the residual $u=1$ renormalon effect,
respectively.
We estimate the uncertainty regarding $1/m_b^2$ by adding 
the ${\cal O}(\LMS^3/m_b^3)$ term of the form $d^{(0)}\rho_{LS}^3/\big[m_b\big]_{\rm PV}^3$
with $d^{(0)}\approx-17$ and $\rho_{LS}^3=2\LMS^3\approx2\times(300~{\rm MeV})^3$.\footnote{
In Ref.~\cite{Alberti:2014yda},  $\rho_{LS}^3$ is determined by the global fit and its size is given by $|\rho_{LS}^3|=0.15~{\rm GeV}^3\gg\LMS^3\approx0.027~\rm GeV^3$.
In our estimation, we assume the natural size of the non-perturbative matrix elements of ${\cal O}(\LMS^n)$
because we do not introduce a factorization scale $\mu_f (\gg \LMS)$.
We multiply the factor $2$ trying to avoid underestimating the effects.
}
In the second line we combine the uncertainties.
While the perturbative uncertainty is at the percent level, 
the uncertainty from $\mbar_b$ is dominant among the systematic uncertainties, which reflects the fact that $\Gamma$ is proportional to the square of $|V_{cb}|$ and the fifth power of $\mbar_b$.
That is, the relative uncertainty of $|V_{cb}|$ and that of $\mbar_b$ are related by
\be
\frac{\delta |V_{cb}|}{|V_{cb}|}\approx-\frac{5}{2}\times\frac{\delta\mbar_b}{\mbar_b}.
\ee
Therefore, it is important to determine $\mbar_b$ accurately 
in order to improve the accuracy of $|V_{cb}|$ determination.

\begin{figure}[t]
\centering
\includegraphics[width=14cm]{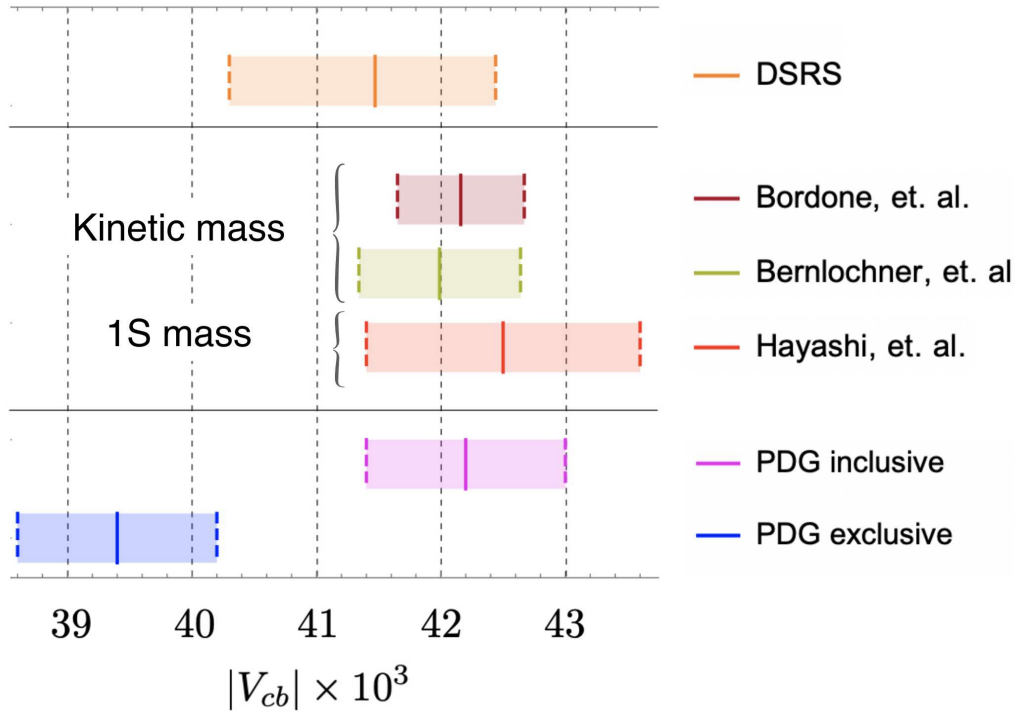}
\caption{\small
Comparison of $|V_{cb}|$ determinations by the DSRS method and previous studies.
}
\label{comp-Vcb}
\end{figure}

\subsection{Comparison with determinations in other mass schemes}

We compare the results of existing $|V_{cb}|$ determinations 
in different short-distance mass schemes; the kinetic mass and 1S mass schemes.
The kinetic mass is defined by introducing a factorization scale $\mu_f$, 
which enables subtraction of the $u=1/2$ and $u=1$ renormalons of 
the quark pole mass simultaneously.
By construction,
the kinetic quark mass and non-perturbative parameters have factorization scale dependence.
Using the latest perturbative calculation of $C^\Gamma_{\bar{Q}Q}$ 
and pole-kinetic mass relation up to ${\cal O}(\alpha_s^3)$, 
$|V_{cb}|$ is determined as  \cite{Bordone:2021oof}
\be
|V_{cb}|=\sqrt{\frac{{\cal B}(B\to X_c\ell \nu)}{(10.66\pm0.15)\%}}\times0.04216(51),
\label{kinetic-Gamma}
\ee
in which ${\cal B}(B\to X_c\ell \nu)$ is the semileptonic branching ratio and the N$^3$LO calculations of the total decay width $\Gamma$ and the NNLO calculation of the lepton energy and hadronic mass moments $d\Gamma/dq^2$ are used.
Another (more recent) 
result using the N$^3$LO calculations of the decay width and the lepton energy moments
is given by~\cite{Bernlochner:2022ucr} 
\bea
|V_{cb}|
&=&\sqrt{\frac{{\cal B}(B\to X_c\ell \nu)}{(10.63\pm0.19)\%}}\times0.04199(65).
\label{kinetic-mom}
\eea

The 1S mass \cite{Hoang:1998hm,Hoang:1998ng} 
is another short-distance mass defined by half of the (perturbatively calculated) mass 
of the heavy quarkonium (a bound state of a heavy quark and an anti-heavy quark).
Since the heavy quarkonium is a sufficiently UV object with a small radius, 
contributions from IR degrees of freedom are decoupled naturally.
$|V_{cb}|$ is determined
using the N$^3$LO perturbative expansion of the 
pole-1S mass relation from the energy levels of two different bottomonium 
states $\Upsilon(1S)$ and $\eta_b(1S)$ \cite{Hayashi:2022hjk}.
The results are given by
\be
|V_{cb}|=0.0421\,(7)_{\rm sys}~~({\rm from}~\Upsilon({\rm 1S})),~~~~
|V_{cb}|=0.0429\,(7)_{\rm sys}~~({\rm from}~\eta_b({\rm 1S})).
\label{Vcb-Up-eta}
\ee
In this determination, the branching ratio ${\cal B}=(10.65\pm0.16)\%$ from the PDG \cite{ParticleDataGroup:2022pth} is used.
Since there is a large difference of the central values in Eq.~\eqref{Vcb-Up-eta},
the combined result~\cite{Hayashi:2022hjk} 
\be
|V_{cb}|=0.0425\,(7)_{\rm sys}\,(8)_{\rm spin}=0.0425(11)
\label{Vcb-spinave}
\ee 
includes a large systematic uncertainty from the difference of the spin 
dependence of the 1S bottomonium states.
This result is free from the $u=1/2$ renormalon but not from
the $u=1$ renormalon.

Fig.~\ref{comp-Vcb} shows the comparison of the results of the DSRS method to
the previous studies.
Each solid line and band width represent the central value and combined uncertainty, respectively.
The first result is ours in the $\msbar$ mass scheme, given by Eq.~\eqref{res-Vcb-PDG}.
The second \cite{Bordone:2021oof} and third \cite{Bernlochner:2022ucr} results are the N$^3$LO results in the kinetic mass scheme, given by 
Eqs.~\eqref{kinetic-Gamma}~and~\eqref{kinetic-mom}, respectively.
The fourth one \cite{Hayashi:2022hjk} is the N$^3$LO result in the 1S mass scheme, given by Eq.~\eqref{Vcb-spinave}.
We note that the results of Refs.~\cite{Bordone:2021oof} and \cite{Bernlochner:2022ucr}
are obtained with global fits using moments, where consistency checks of the OPE are therefore provided,
while our present study and Ref.~\cite{Hayashi:2022hjk} use only the total decay width.
The last two values are the PDG inclusive and exclusive values given by Eq.~\eqref{Vcb-PDG}.
We note that these determinations use the similar values for ${\cal B}$.\footnote{
Recently some experimental data have been reported, which indicate
a smaller value for ${\cal B}$.
}
Our result in the $\msbar$ mass scheme is consistent with
the results in other mass schemes within the uncertainty.
It is also consistent with the PDG value from the inclusive decays, while
it is in tension with that from the exclusive decays.

\begin{figure}[t]
\centering
\includegraphics[width=10cm]{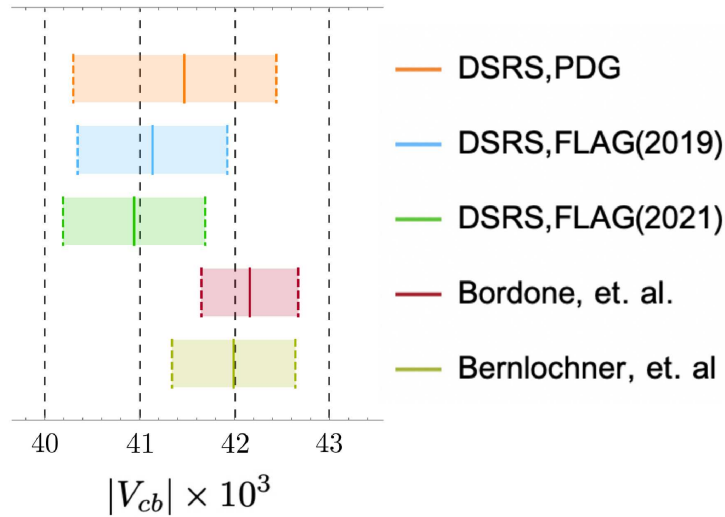}
\caption{\small
Comparison of $|V_{cb}|$ with various input values of heavy quark masses.
}
\label{comp-Vcb-kin}
\end{figure}

Now we discuss the size of the uncertainty of our result in comparison
to those of the kinetic mass schemes, which are smaller by a factor of
two or so.
As already noted, the largest source of the uncertainty of our result
stems from the uncertainty of the input bottom 
quark $\msbar$ mass.
Furthermore, the uncertainty of the charm quark $\msbar$ mass
is one of the large uncertainties.
We use the PDG values for these input masses.
On the other hand, in the kinetic mass scheme determinations,
the more accurate input $\msbar$ masses,
$\mbar_b(\mbar_b)=4.198\pm0.012~{\rm GeV}$ and  $\mbar_c(\mbar_c)=1.280\pm0.013~{\rm GeV}$ taken from the FLAG2019 \cite{FlavourLatticeAveragingGroup:2019iem}, are used.
For comparison, the $|V_{cb}|$ value determined in our method
by using the quark masses from the FLAG2019 is given by
\bea
|V_{cb}|
&=&0.0411\,(4)_{\rm PT}\,(4)_{\mbar_b}\,(2)_{\mbar_c}\,(3)_{\alpha_s}\,(1)_{\tau_B}\,(4)_{\cal B}\,(1)_{\mu_\pi^2}\,(0)_{\mu_G^2}\,(0)_{{\rm h.o.}\,C_{cm}}\,(1)_{1/m_b^3}\,(0)_{{\rm sub}\,u=1}\non
&=&0.0411\,(8),
\label{res-Vcb-FLAG19}
\eea
where the convention of the uncertainties is the same as Eq.~\eqref{res-Vcb-PDG}.
Fig.~\ref{comp-Vcb-kin} shows a comparison of $|V_{cb}|$ with several input quark masses.
We add the result with the FLAG 2021 masses of the bottom and charm quarks as in Eqs.~\eqref{FLAG2021mb} and \eqref{FLAG2021mc}.
We see that
the uncertainty size becomes more or less
comparable to the more recent
kinetic-mass scheme result in Eq.~\eqref{kinetic-mom}.
Nevertheless, given the current status of the FLAG results
with different numbers of dynamical quarks (see Eqs.~\eqref{FLAG2021mb0}
and \eqref{FLAG2021mb}), we choose to take a conservative
attitude to use the PDG masses.

\section{Summary and Conclusions}
\label{chap5}

As one of the precision tests of the SM, we performed
a determination of $|V_{cb}|$ from the inclusive semileptonic decay width.
We employed the $\msbar$ mass scheme to examine
theoretical consistency of inclusive determination of $|V_{cb}|$,
which has been performed in other mass schemes.
The LO Wilson coefficient in the OPE of $\Gamma(B\to X_c\ell\nu)$ 
in the $\msbar$ mass scheme has the $u=1$ renormalon, 
which is absorbed by the first non-perturbative 
matrix element
$\mu_\pi^2$ (one of the HQET parameters). 
We use the DSRS method to remove the renormalon from the Wilson coefficient
and to determine $\mu_\pi^2$ with renormalon subtraction.

$\mu_\pi^2$ is determined from the masses of the $B$ and $D$ mesons.
The LO term of the $1/m_h$ expansion of the heavy-light
meson mass is the heavy quark pole mass $m_h$.
HQET describes the non-perturbative corrections
in a systematic expansion in $1/m_h$.
From the quark pole masses $m_b$ and $m_c$, we separated the $u=1/2$ and $u=1$ renormalons, which are canceled by the HQET parameters $\bar{\Lambda}$ and $\mu_\pi^2$, respectively.
The HQET parameters $\big[\bar{\Lambda}\big]_{\rm PV}$ and $\big[\mu_\pi^2\big]_{\rm PV}$, which are defined in the infinite mass limit and with renormalons removed, are important parameters used to predict multiple observables of the $B$ and $D$ mesons. 
Using the DSRS method for the known 
pole-$\msbar$ mass relation up to ${\cal O}(\alpha_s^4)$ (and with estimated $\alpha_s^5$ coefficient) with the PDG values of $\mbar_b$ and $\mbar_c$ as inputs, 
the renormalon-subtracted quark pole masses
in the PV scheme, called the PV masses, are determined as
\be
\big[m_b\big]_{\rm PV}=4.822\,(36)~{\rm GeV},\quad\big[m_c\big]_{\rm PV}=1.468\,(35)~{\rm GeV},
\ee
where the uncertainties represent combined systematic uncertainties.
The perturbative series show expected good convergent behaviors, and
both PV masses would
have smaller theoretical uncertainties when perturbative calculations of 
the next order are achieved. 
This is owing to the removal of the renormalons. 
On the other hand, in each mass determination, the uncertainty due to the input value of the $\msbar$ mass
is large.
More precise determination of the $\msbar$ masses is an important task for future precision physics.

Using the results of the PV masses, we determined $\big[\bar{\Lambda}\big]_{\rm PV}$ and $\big[\mu_\pi^2\big]_{\rm PV}$ as
\be
\big[\bar{\Lambda}\big]_{\rm PV}=0.486\,(54)~{\rm GeV},\quad\big[\mu_\pi^2\big]_{\rm PV}=0.05\,(22)~{\rm GeV^2},
\ee 
where the systematic uncertainties are combined.
The size of the (combined) systematic uncertainty for $\big[\bar{\Lambda}\big]_{\rm PV}$ is sufficiently small, reflecting the fact that the IR renormalons of the pole mass are properly removed.
The $\big[\mu_\pi^2\big]_{\rm PV}$ result has apparently large perturbative uncertainty, but this would be due to the lack of higher-order
perturbative coefficients in the pole-$\msbar$ mass relation (especially for the charm quark).
The results are consistent with the previous studies in the same 
(PV) scheme \cite{FermilabLattice:2018est,Ayala:2019hkn}.

The DSRS method is also applied to the LO Wilson coefficient of the semileptonic $B$ decay width 
$\Gamma(B\to X_c\ell \nu)$.
Although the $\msbar$ mass is not favored in previous studies 
due to the large perturbative corrections between the pole mass and the $\msbar$ mass, 
we found that the convergence of the Wilson coefficient improves when the DSRS method is applied;
the DSRS method can suppress the $u=1$ renormalon 
and eliminate the unphysical singularity at $\mu\sim\LMS$.
The DSRS result is consistent with the fixed-order calculation which does not subtract the $u=1$ renormalon, and has smaller scale dependence.

Our result of $|V_{cb}|$ determination using the renormalon-subtracted Wilson coefficient is given by
\be
|V_{cb}|=0.0415\,(^{+10}_{-12}),
\ee
where the uncertainties are combined.
We use the PDG values for the input parameters $\mbar_b$, $\mbar_c$, and $\alpha_s$. 
We incorporate the non-perturbative corrections described by two HQET parameters $\big[\mu_\pi^2\big]_{\rm PV}$ and $\mu_G^2$, 
which are determined from the masses of 
the $B$ and $D$ mesons, and from the hyperfine-splitting of the $B$ mesons, respectively.
The experimental input values of the branching ratio 
and the lifetime are close to the ones used in the previous studies.
The uncertainty due to the perturbative calculation is reduced to an accuracy of one percent, due to a series with good convergence constructed using the DSRS method.
On the other hand, the uncertainty from the input parameters, 
in particular the bottom quark mass, 
is large.
(Our error estimate is based on the PDG bottom and charm mass values.)

Our results for $\big[\bar{\Lambda}\big]_{\rm PV}$, 
$\big[\mu_\pi^2\big]_{\rm PV}$ and $|V_{cb}|$
are consistent with those of the previous studies using 
other short-distance mass schemes or other renormalon subtraction
schemes.
In particular, the value as well as the uncertainty of
$|V_{cb}|$ are consistent with other
determinations from the inclusive decays, 
provided uncertainties of the input bottom and charm masses
are appropriately taken into account.
The determined $|V_{cb}|$ value has a tension with
that from the exclusive decays.
Thus, our study provides a consistency check
of the theoretical calculations used in $|V_{cb}|$ determination from the inclusive
decays.

\section*{Acknowledgements}
Y.H. acknowledges support from GP-PU at Tohoku University. 
This work was supported by JSPS KAKENHI Grant Numbers
JP21J10226, JP20J00328, JP20K03923, JP19H00689, JP19K14711, 
and JP18H05542.

\appendix
\section{Perturbative coefficients}
\label{App.A}
We collect the perturbative coefficients
necessary for the analyses in this paper.
\subsection*{QCD beta function}
The QCD $\beta$ function is known up to ${\cal O}(\alpha_s^6)$ (5-loop accuracy)
\cite{Baikov:2016tgj,Baikov:2017ayn,Herzog:2017ohr, Luthe:2017ttg}:
\be
\beta(\alpha_s)=-\sum_{i=0}^4b_i\alpha_s^{i+2},
\ee
\be
b_0=\frac{1}{4\pi}\Big(11-\frac{2}{3}n_f\Big),\quad
b_1=\frac{1}{(4\pi)^2}\Big(102-\frac{38}{3}n_f\Big),
\ee
\be
b_2=\frac{1}{(4\pi)^3}\Big(\frac{2857}{2}-\frac{5033}{18}n_f+\frac{325}{54}n_f^2\Big),
\ee
\bea
b_3=\frac{1}{(4\pi)^4}\Big[\frac{149753}{6}
&+&3564\zeta_3-\Bigg(\frac{1078361}{162}+\frac{6508}{27}\zeta_3\Big)n_f\nonumber\\
&+&\Big(\frac{50065}{162}+\frac{6472}{81}\zeta_3\Big)n_f^2+\frac{1093}{729}n_f^3\Bigg],
\eea
\bea
b_{4}&=&\frac{1}{(4\pi)^{5}}\Bigg[\frac{8157455}{16}+\frac{621885}{2} \zeta_{3}-\frac{88209}{2} \zeta_{4}-288090 \zeta_{5}\nonumber\\
&+&n_{f}\left(-\frac{336460813}{1944}-\frac{4811164}{81} \zeta_{3}+\frac{33935}{6} \zeta_{4}+\frac{1358995}{27} \zeta_{5}\right)\nonumber\\
&+&n_{f}^{2}\left(\frac{25960913}{1944}+\frac{698531}{81} \zeta_{3}-\frac{10526}{9} \zeta_{4}-\frac{381760}{81} \zeta_{5}\right)\nonumber\\
&+&n_{f}^{3}\left(-\frac{630559}{5832}-\frac{48722}{243} \zeta_{3}+\frac{1618}{27} \zeta_{4}+\frac{460}{9} \zeta_{5}\right)+n_{f}^{4}\left(\frac{1205}{2916}-\frac{152}{81} \zeta_{3}\right)\Bigg].
\eea
$n_f$ is the number of active quark flavors, and $\zeta_n=\sum_{k=1}^\infty k^{-n}$
is the Riemann zeta function.

\subsection*{Mass of heavy-light meson}
Based on HQET,
the $1/m_h$ expansion of the heavy-light meson whose mass is $M_H$ is given by
\be
M_H^{(s)}=m_h+\bar{\Lambda}+\frac{\mu_\pi^2}{2m_h}+A(s)C_{cm}(m_h)\frac{\mu_G^2(m_h)}{2m_h}+{\cal O}\bigg(\frac{\LMS^3}{m_h^2}\bigg),
\ee
where $s\,(=0,\,1)$ denotes the spin of $H$.
The Wilson coefficient of the chromo-magnetic term $C_{cm}$ 
is calculated up to ${\cal O}(\alpha_s^3)$ \cite{Grozin:2007fh}, which was used to evaluate $\mu_G^2(\mbar_b)$ \cite{Hayashi:2022hjk} .
The numerical values of the coefficients of
\be
C_{cm}=1+\sum_{n=0}^2\alpha_s(m_h)^{n+1}c^{cm}_n,
\ee
with $\alpha_s=\alpha_s^{(n_f)}$, are given by
\be
c^{cm}_0\approx 0.6897,
\ee
\be
c^{cm}_1\approx 2.2186-0.1938\, n_f,
\ee
and
\be
c^{cm}_2\approx 11.079-1.7490\, n_f+0.0513 \,n_f^2.
\ee
\subsection*{Total decay width of inclusive semileptonic $B$ decay}
The OPE of the total decay width of $B\to X_c\ell\nu$ is given by
\be
\Gamma=\frac{G_F^2|V_{cb}|^2}{192\pi^3}A_{ew}m_b^5\bigg[C_{\bar{Q}Q}^\Gamma\Big(1-\frac{\mu_\pi^2}{2m_b^2}\Big)+C_{cm}^\Gamma\frac{\mu_G^2}{m_b^2}+{\cal O}\bigg(\frac{\LMS^3}{m_b^3}\bigg)\bigg].
\ee
The leading Wilson coefficient $C_{\bar{Q}Q}^\Gamma$ in terms of a pole mass is perturbatively calculated as
\be
C_{\bar{Q}Q}^\Gamma=\sum_{n=0}^3\alpha_s(m_b^2)^nX_n(\rho),
\ee
with $\rho=m_c/m_b$ and $\alpha_s=\alpha_s^{(5)}$.
$X_0$, $X_1$, and $X_2$ are calculated \cite{Luke:1994yc,Trott:2004xc,Aquila:2005hq,Pak:2008qt,Pak:2008cp,Melnikov:2008qs,Dowling:2008mc}
as
\be
X_0=1-8\rho^2-12\rho^4\log(\rho^2)+8\rho^6-\rho^8,
\ee
\bea
X_1
&=&-\frac{2}{3 \pi }\bigg[-(1-\rho^4)\Big(\frac{25}{4}-\frac{239}{3}\rho^2+\frac{25}{4}\rho^4\Big)+\rho^2\log(\rho^2)\Big(20+90\rho^2-\frac{4}{3}\rho^4+\frac{17}{3}\rho^6\Big)\non
&&~~~~~~~~
+\rho^4 \log^2(\rho^2) (36 + \rho^4)+(1-\rho^4)\Big(\frac{17}{3}-\frac{64}{3}\rho^2+\frac{17}{3}\rho^4\Big)\log(1-\rho^2)\non
&&~~~~~~~~
-4(1+30\rho^4+\rho^8)\log(\rho^2)\log(1-\rho^2)-(1+16\rho^4+\rho^8)(6{\rm Li}_2(\rho^2)-\pi^2)\non
&&~~~~~~~~
-32\rho^3(1+\rho^2)(\pi^2 - 4{\rm Li}_2(\rho)+ 4 {\rm Li}_2(-\rho)- 
  2 \log(\rho^2) \log\bigg(\frac{1 - \rho}{1 + \rho}\bigg)
\bigg],
\eea
and $X_2$ is known in the expansion in $\rho$,
\bea
X_2&\approx&-2.158 - 0.8333 \rho + (-65.01- 
 39.22 \log(\rho) + 0.2701 \log^2(\rho) ) \rho^2 \non
&& +(- 118.7 - 129.8 \log(\rho))\rho^3 + 
 (128.2- 124.6 \log(\rho)- 16.52 \log^2(\rho)+ 1.081 \log^3(\rho)) \rho^4\non
 &&( - 41.65   - 80.98 \log(\rho) )\rho^5 
 + (98.42 - 39.30 \log(\rho)+ 16.77 \log^2(\rho))\rho^6\non
 &&(- 14.86 + 1.954 \log(\rho))\rho^7+ (0.09796-  0.1094 \log(\rho))\rho^8 +{\cal O}(\rho^9) .
\eea
$X_3$ is known as the expansion in $\delta=1-\rho$ up to ${\cal O}(\delta^{20})$ \cite{Fael:2020tow}.

$C_{cm}^\Gamma$ has been calculated to ${\cal O}(\alpha_s)$ order \cite{Bigi:1992su,Blok:1993va,Manohar:1993qn,Alberti:2013kxa}.
In this paper, to determine the central value of $|V_{cb}|$, we use the LO term $C_{cm}^\Gamma\big|_{\rm LO}$ given by
\be
C_{cm}^\Gamma\big|_{\rm LO}=-\frac{1}{2}(3-8\rho^2+24\rho^4 -24\rho^6 +5\rho^8 +12\rho^4\log(\rho^2)).
\label{CcmLO}
\ee
The NLO term is used for the estimation of uncertainty, which is given by
\be
C_{cm}^\Gamma\big|_{\rm NLO}\approx-2.42\times X_0(\bar{\rho})\alpha_s(\mbar_b),
\label{CcmNLO}
\ee
where $``\approx"$ means that we use the numerical value of $C_{cm}^\Gamma\big|_{\rm NLO}$ at $\mu=m_b$ since the analytic result is not available.

To change the flavor of running coupling constant from $n_f$ flavors to $(n_f-1)$ flavors,
we use a flavor threshold relation \cite{Chetyrkin:1997sg} given by
\bea
\frac{\alpha_s^{(n_f-1)}}{\alpha_s^{(n_f)}}
&=&1-\frac{\ell_h}{6\pi}\alpha_s^{(n_f)}+\bigg(\frac{\ell_h^2}{36}-\frac{19}{24}\ell_h+c_2\bigg)\big(\alpha_s^{(n_f)}\big)^2\non
&&\!\!+\bigg(-\frac{\ell_h^3}{216}-\frac{131}{576}\ell_h^2+\frac{-6793+281(n_f-1)}{1728}\ell_h+c_3\bigg)\big(\alpha_s^{(n_f)}\big)^3+{\cal O}(\alpha_s^4),\,\,
\label{flth}
\eea
with $\alpha_s=\alpha_s(\mu^2)$, $\mu_h=\mbar(\mu_h)$, $\ell_h=\log(\mu^2/\mu_h^2)$ and
\be
c_2=\frac{11}{72},\quad c_3=-\frac{82043}{27648}\zeta(3)+\frac{564731}{124416}-\frac{2633}{31104}(n_f-1).\,\,
\ee
In this paper, we take a scale for matching $\alpha_s^{(5)}$ to $\alpha_s^{(4)}$ as $\mu=\mu_h=\mbar_b$, and  for matching $\alpha_s^{(4)}$ to $\alpha_s^{(3)}$ as $\mu=\mu_h=\mbar_c$.

We note that in the matching relation between $\alpha_s^{(5)}$ and $\alpha_s^{(3)}$, power terms in $m_c/m_b$ appear in the third order coefficient \cite{Grozin:2011nk}
(which does not appear in the two-step matching using the above matching relation).
In our study, this effect is relevant in the procedure to obtain the three-flavor $\LMS$ from $\alpha_s^{(5)}(M_Z^2)$.
Based on the discussion given in Ref.~\cite{Grozin:2011nk}, we expect that the effect of these terms is smaller than our uncertainty and therefore neglect them.
We also note that the power terms are not relevant to obtaining the perturbative series in the three-flavor coupling within our study.
When we give the perturbative series for the pole-$\overline{\text{MS}}$ mass relation in the three-flavor coupling,
this effect is relevant to the $\alpha_s^4$ order term and beyond but we use estimated coefficients at these orders.
The perturbative series for the semileptonic decay width is given to the $\alpha_s^3$ order,
where the power terms are irrelevant.

\section{\boldmath Estimate of $d_4^{(h)}$}
\label{App.B}

We estimate $d_4^{(h)}$ based on the asymptotic form of 
the perturbative coefficients governed by the $u=1/2$ renormalon. (See, e.g., Eq.~(30) of Ref.~\cite{Takaura:2021yaj} with $r$ replaced by $\mbar_h$.) 
The only unknown parameter in the asymptotic form is the normalization constant,
which can be estimated by the ratio of the actual coefficients (up to $d^{(h)}_3$) and the coefficients of 
the asymptotic form.
We give the central value of the normalization constant by $d^{(h)}_3/d^{(h) ({\rm asym})}_3$.
$d^{(h) ({\rm asym})}_n$ is obtained in the expansion in $1/n$ and 
we include up to $\mathcal{O}(1/n^3)$ (which is possible with the knowledge of $b_4$)
in giving $d^{(h) ({\rm asym})}_3$.
The uncertainty of the normalization constant is estimated by the difference
between $d^{(h)}_2/d^{(h) ({\rm asym})}_2$ and $d^{(h)}_3/d^{(h) ({\rm asym})}_3$.
We confirmed that 
when we give $d^{(h) ({\rm asym})}_3$ up to the $\mathcal{O}(1/n^2)$ term for 
$d^{(h)}_3/d^{(h) ({\rm asym})}_3$,
the variation of the normalization constant is smaller than the uncertainty determined above.
Using this normalization constant and its uncertainty, we obtain Eq.~\eqref{d4est}.
There are previous works estimating $d_4^{(h)}$ using similar and different methods. 
See, e.g., Refs.~\cite{Ayala:2014yxa,Kataev:2019zfx}.


\bibliographystyle{JHEP}

\providecommand{\href}[2]{#2}\begingroup\raggedright\endgroup

\end{document}